\documentclass[fleqn,usenatbib]{mnras}
\usepackage{newtxtext,newtxmath}
\usepackage[T1]{fontenc}
\usepackage{ae,aecompl}
\usepackage{multicol,multirow}
\usepackage{graphicx,stmaryrd}	% Including figure files
\usepackage{amsmath}	% Advanced maths commands
\usepackage{amssymb}	% Extra maths symbols
\usepackage{mathtools}
\usepackage{soul}
\usepackage[normalem]{ulem}
\title[Observing Dust Trapping in SG Discs]{The observational impact of dust trapping in self-gravitating discs}

\author[J. Cadman et al.]{\parbox{\textwidth}
{James Cadman$^{1,2},$\thanks{E-mail: \texttt{cadman@roe.ac.uk}}
Cassandra Hall$^{3,4,5},$
Ken Rice$^{1,2},$
Tim J. Harries$^{6}$ and \\
 Pamela D. Klaassen$^{7}$}\vspace{0.4cm}
\\
$^{1}$SUPA, Institute for Astronomy, University of Edinburgh, Blackford Hill, Edinburgh, EH9 3HJ, Scotland, UK\\
$^{2}$Centre for Exoplanet Science, University of Edinburgh, Edinburgh, UK\\
$^{3}$Department of Physics and Astronomy, University of Leicester, University Road, Leicester, LE1 7RH, UK\\
$^{4}$Department of Physics and Astronomy, The University of Georgia, Athens, GA 30602, USA. \\
$^{5}$Center for Simulational Physics, The University of Georgia, Athens, GA 30602, USA.\\ 
$^{6}$Department of Physics and Astronomy, University of Exeter, Stocker Road, Exeter, EX4 4QL, UK\\
$^{7}$UK Astronomy Technology Center, Royal Observatory Edinburgh, Blackford Hill, Edinburgh, EH9 3HJ, UK\\
}

\date{Accepted 2020 August 21. Received 2020 August 21; in original form 2020 May 26
}

\pubyear{2020}

\begin{document}
\label{firstpage}
%\pagerange{\pageref{firstpage}--\pageref{lastpage}}
\maketitle 

% Abstract of the paper
\begin{abstract}
We present a 3D semi-analytic model of self-gravitating discs, and include a prescription for dust trapping in the disc spiral arms. Using Monte-Carlo radiative transfer we produce synthetic ALMA observations of these discs. In doing so we demonstrate that our model is capable of producing observational predictions, and able to model real image data of potentially self-gravitating discs. For a disc to generate spiral structure that would be observable with ALMA requires that the disc's dust mass budget is dominated by millimetre and centimetre-sized grains. Discs in which grains have grown to the grain fragmentation threshold may satisfy this criterion, thus we predict that signatures of gravitational instability may be detectable in discs of lower mass than has previously been suggested. For example, we find that discs with disc-to-star mass ratios as low as $0.10$ are capable of driving observable spiral arms. Substructure becomes challenging to detect in discs where no grain growth has occurred or in which grain growth has proceeded well beyond the grain fragmentation threshold. We demonstrate how we can use our model to retrieve information about dust trapping and grain growth through multi-wavelength observations of discs, and using estimates of the opacity spectral index. Applying our disc model to the Elias 27, WaOph 6 and IM Lup systems we find gravitational instability to be a plausible explanation for the observed substructure in all 3 discs, if sufficient grain growth has indeed occurred.
\end{abstract}

% Select between one and six entries from the list of approved keywords.
% Don't make up new ones.

%%%%%%%%%%%%%%%%%%%%%%%%%%%%%%%%%%%%%%%%%%%%%%%%%%

%%%%%%%%%%%%%%%%% BODY OF PAPER %%%%%%%%%%%%%%%%%%

\section{Introduction}\label{sec:intro}

Discs around very young stars are typically heavily embedded and optically thick to optical wavelengths \citep{dunhametal14}.  They will, however, emit thermal infrared (IR) radiation and may be resolved by high-resolution, sub-mm observations with the Atacama Large Millimeter/submillimeter Array (ALMA). Thanks to recent observational advances, spiral substructure, characteristic of massive self-gravitating protoplanetary discs, is now within our observing capabilities. \citep{perezetal16,dsharp1,dsharp2}.

Non-axisymmetric structure will manifest as spiral density perturbations when \citep{durisenetal07},
\begin{equation}\label{eq:Q}
    Q=\frac{c_{\rm s}\kappa}{\pi \mathrm{G} \Sigma} \lesssim 1.5-1.7,
\end{equation}
where $Q$ is the Toomre parameter \citep{toomre64}, $c_{\rm s}$ is the disc sound speed, $\kappa$ is the epicyclic frequency (equal to the angular frequency, $\Omega$, in a Keplerian disc), G is the gravitational constant and $\Sigma$ is the disc surface density. 

From inspection of $Q$, it is clear that more massive discs (higher $\Sigma$) will be susceptible to gravitational instabilities (hereafter GI), and it is likely that in the earliest stages of a protoplanetary disc's lifetime they may be massive enough to generate prominent spiral structure \citep{linpringle87,linpringle90,riceetal10}.  The mass accretion rate in these massive discs is likely to be high \citep{riceetal10} and, hence, once a disc is no longer being replenished by envelope infall, it will be rapidly depleted. Consequently, unless the envelope is optically thin at the relevant observing wavelengths, signatures of GI will only be detectable for about $10^4$ years after accretion through the disc begins to dynamically dominate over infall from the envelope \citep{halletal19}.

The Disk Substructures at High Angular Resolution Project (DSHARP) ALMA survey recently performed an in depth analysis of 20 nearby protoplanetary discs, 3 of which exhibit possible spiral substructure reminiscent of GI \citep{perezetal16,dsharp1,dsharp3}. 

Non-axisymmetric disc features are not unique to GI, and may be explained through alternative mechanisms such as planet-disc interactions \citep{linpapaloizou86,tanakaetal02}. It may be possible to distinguish between planet and GI induced spiral structure through scattered light vs. sub-mm observations, as dust trapping in spiral regions is likely to be more effective in gravitationally unstable discs \citep{riceetal04,dongetal15b,juhaszetal15}.

Spiral density perturbations in self-gravitating discs act as pressure traps for dust grains, which will radially migrate and concentrate at the pressure maxima \citep{riceetal04}. Due to the negative outward gas pressure gradient, in a smooth, laminar disc, gas particles orbit with slightly sub-Keplerian velocities compared to solids at the same radii. Since the outward gas pressure gradient doesn't directly influence the solids, this can produce a significant gas drag on the faster orbiting dust grains, resulting in their radial migration. Micron-sized grains, however, will typically be strongly coupled to the gas, hence will orbit with the same, sub-Keplerian velocities and will closely trace the gas distribution. Metre-sized, and larger, objects will be largely decoupled and will orbit with approximately Keplerian velocities. Intermediate, $\sim$mm-sized grains will however experience a large radial drift. 

In smooth, laminar discs radial drift results in migration toward the disc centre where gas pressure is maximum. However, the propagation of GI induced spiral density perturbations will generate a non-axisymmetric pressure gradient, resulting in significant concentration of mm-sized grains at the peaks of the spiral density waves. This will have important consequences; producing enhanced emission in these regions as well as potentially accelerating planetesimal growth \citep{riceetal04,riceetal06}. \cite{dipierro14,dipierro15} have previously shown that GI induced spiral structure should be detectable with ALMA at moderate distances ($d\sim140$\,pc), and that dust migration as a result of self-gravitating disc structure will produce detectable signatures in their observed spectral index maps.

In this paper we build on previous work by \cite{halletal16} who developed a semi-analytic formalism for determining the structure of self-gravitating protostellar discs, performed 3D Monte Carlo radiative transfer on these models and produced synthetic disc images using the ALMA simulator.  We add to this by including a prescription for the effects of dust grain enhancement in the spiral density waves. These models allow us to produce a suite of discs at little computational expense when compared to approaches such as Smoothed Particle Hydrodynamics (SPH). Therefore, we are able to efficiently explore a wide range of disc parameter space and produce observational predictions for telescopes such as ALMA.

In Sections \ref{sec:discmodel} and \ref{sec:casa} we present our disc model setup, and describe the radiative transfer approach as well as how we used the ALMA simulator in our analysis. In Section \ref{sec:sphdust} we use SPH to model the extent to which we might expect grains to be enhanced in self-gravitating discs, allowing us to inform our semi-analytic prescription. In Section \ref{sec:grainsizes} we discuss grain growth and the fragmentation threshold. In Section \ref{sec:discparams} we discuss our disc parameter setup and in Section \ref{sec:taurusdiscresults} we apply our disc models to discs comparable to those in the Taurus star-forming region, presenting observational predictions for observing self-gravitating discs at distance $\sim140$\,pc. In Section \ref{sec:dsharp} we apply our models to three discs from the DSHARP survey, analysing whether or not their observed substructure may be the result of self-gravity. In Section \ref{sec:discussion} we discuss and draw conclusions.

\section{Disc Models: Setup}\label{sec:discmodel}

We setup our discs using the 1D models introduced by \cite{clarkeetal09} \citep[see also][]{ricearmitage09, forganrice2013} and further developed by \cite{halletal16} to include 3D structure such as the spiral density waves characteristic of self-gravitating discs. These models are described in detail in \cite{halletal16} and summarised in Section \ref{sec:sgdiscmodels}.
We refer the reader to \cite{halletal16} for a comparison of this simple functional formailsm's ability to accurately reproduce self-gravitating spiral shape and amplitudes from SPH simulations. Dust grain enhancement is imposed semi-analytically, in line with what we might expect from spiral density structure in self-gravitating discs, and is described in Section \ref{sec:dustconc}.

\subsection{Self-gravitating Disc Models}\label{sec:sgdiscmodels}
We expect an accretion disc to settle into a quasi-steady state \citep{paczynski78,gammie01,ricearmitage09} with a constant mass accretion rate, $\dot{M}$, given by \citep{pringle81},
\begin{equation}\label{eq:mdot}
    \dot{M} = \frac{3\pi\alpha c_s^2\Sigma}{\Omega} = \rm constant,
\end{equation}
where $c_s$ is the local sound speed, $\Sigma$ is the disc surface density, $\alpha$ is the 
dimensionless viscosity parameter \citep{shakurasunyaev73}, and $\Omega$ is the Keplerian angular frequency.  Strictly speaking, a self-gravitating disc is not actually viscous, but the stresses can still be represented by an effective viscous-$\alpha$ parameter \citep{balbus99,gammie01,lodato04}. Assuming local angular momentum transport, and that the disc is in thermal equilibrium, this can be expressed as \citep{gammie01},
\begin{equation}\label{eq:alpha}
    \alpha = \frac{4}{9\gamma(\gamma - 1)t_{\rm cool}\Omega},
\end{equation}
where $\gamma$ is the ratio of specific heats and $t_{\rm cool}$ is the local cooling timescale.

Cooling is modelled in terms of a local cooling rate, $\Lambda$. In the presence of external irradiation that we express as a temperature, $T_{\rm irr}$, the local cooling rate can be expressed as \citep{hubney90},
\begin{equation}\label{eq:lambda}
    \Lambda = \frac{8\sigma(T^4 - T_{\rm irr}^4)}{3\tau},
\end{equation}
where $\sigma$ is the Stefan-Boltzmann constant, $T$ is the midplane disc temperature and $\tau$ represents the optical depth. For all the models considered here we assume that irradiation leads to a constant background temperature, $T_{\rm irr}=10$\,K. The local cooling timescale is then the thermal energy per unit area divided by this cooling rate, which we can write as,
\begin{equation}\label{eq:tcool}
    t_{\rm cool} = \frac{1}{\Lambda}\frac{c_s^2 \Sigma}{\gamma(\gamma -1)}.
\end{equation}

Disc instability is characterised by the Toomre $Q$ parameter \citep[equation \ref{eq:Q},][]{toomre64}
where a disc will be susceptible to non-axisymmetric perurbations when $Q < 1.5-1.7$ \citep{durisenetal07}. Here we assume the disc to be marginally unstable with $Q=2$ at all radii. We can then use equations  \ref{eq:Q}, \ref{eq:mdot}, \ref{eq:alpha} and \ref{eq:tcool} to self-consistently determine values for $\alpha$, $\Sigma$ and $c_s$. This then allows for calculation of the local scale height, $H = c_s/\Omega$, and the midplane volume density, $\rho = \Sigma/2H$. Values for $T$, $\gamma$ and the local optical depth, $\tau = \Sigma\kappa(\rho,c_s)$, are determined from $\rho$ and $c_s$ using the equation of state from \cite{stamatellosetal07}. Temperature and surface density profiles are thus determined self-consistently in these discs, as for any given $\dot{M}$ and disc size there is only one possible combination of $T$ and $\Sigma$ that will satisfy equations \ref{eq:mdot}$-$\ref{eq:tcool}. In this way we are able to construct 3D axisymmetric discs for any desired $\dot{M}$ and disc size.

We then impose spiral density structure as described in \cite{halletal16}. This is done by assuming logarithmic spirals with azimuthal position,
\begin{equation}\label{eq:spiraltheta}
    \theta_{\rm spiral} = \frac{1}{b} {\rm log}\Big(\frac{r}{a}\Big),
\end{equation}
where $a$ and $b$ are constants defining the shape of the spirals. Here we use $a=13.5$ and $b=0.38$, in line with that used in \cite{halletal16}.

At each azimuthal location in the disc, $\theta_{\rm x,y}$, we calculate a fractional over-density, $\delta\Sigma/\Sigma$, characterised by a spiral amplification factor, $S$, such that \citep{cossinslodatoclarke},
\begin{equation}\label{eq:dsigma/sigma}
    \frac{\langle\delta\Sigma\rangle}{\langle\Sigma\rangle} = S\alpha^{1/2},
\end{equation}
where here we define $S=2$, and $\alpha$ is the effective viscous alpha from Equation \ref{eq:alpha} which is determined self-consistently.

This fractional over-density is imposed sinusoidally at each azimuthal location in the disc, $\theta_{\rm x,y}$ such that,
\begin{equation}\label{eq:dsigmamax}
    \delta\Sigma(\phi)=\langle\delta\Sigma\rangle {\rm cos}(m\phi).
\end{equation}
Here, $m$ is the azimuthal wavenumber (i.e. the number of spiral arms) and $\phi$ is the phase difference between the location of the spiral arms and each azimuthal position in the disc,
\begin{equation}
    \phi=\theta_{\rm spiral} - \theta_{\rm x,y}.
\end{equation}
We expect that the azimuthal wavenumber will be roughly related to the disc-to-star mass ratio, $q$, as \citep{cossinslodatoclarke,dongetal15},
\begin{equation}\label{eq:namrs}
    m \approx 1/q. 
\end{equation}
We use this in equation \ref{eq:dsigmamax} to impose an azimuthal wavenumber in a disc of mass-ratio, $q$, assuming a symmetrical response (with $m=2,4,8...$) and rounding $m$ to the nearest appropriate value.

Finally, we model the vertical density profile of the disc as \citep{spitzer42},
\begin{equation}
    \rho(z)=\rho_0\Bigg[\frac{1}{{\rm cosh}^2\Big(\frac{z}{H_{\rm sg}}\Big)}\Bigg],
\end{equation}
where $H_{\rm sg}$ is the self-gravitating scale height given as,
\begin{equation}
    H_{\rm sg}=\frac{c_s^2}{\pi G\Sigma}.
\end{equation}

\subsection{Grain Concentration}\label{sec:dustconc}

In the presence of spiral density waves, dust grains will radially migrate and concentrate at their density maxima \citep{riceetal04}. The extent of this radial migration will be strongly dependent on grain size, $a$. Small grains of $\sim \rm \mu m$ scale will be strongly coupled to the gas in the disc, will experience very little radial drift and will closely trace the gas distribution. The largest particles of $\sim \rm m$ scale will be decoupled and will be unaffected by the disc gas pressure, therefore orbiting with approximately Keplerian velocities.

For intermediate-sized dust grains of $\sim {\rm mm-cm}$ scale, the impact of the gas drag will be significant. Radial drift velocities will be large and, hence, grain concentration at spiral pressure maxima will be high. The gas-dust coupling is characterised by the Stokes number, 
\begin{equation}
    {\rm St}=\frac{a\rho_{\rm s}\Omega}{\rho c_{\rm s}},
\end{equation}
where $\rho_{\rm s}$ is the internal density of the dust grains and $\rho$ is the local gas density.

The solution of the momentum equation suggests that the radial drift velocity has a $1/({\rm St} + {\rm St^{-1}})$ relation \citep{weidenschilling77}. We therefore propose a grain enhancement factor of the form,
\begin{equation}\label{eq:epsilon}
    \eta_i = 1 + \frac{2d}{{\rm St}_i + {\rm St}_i^{-1}} - \frac{{\rm St}_i}{200},
\end{equation}
where $d$ is a constant, to be determined later, that represents the peak dust concentration factor in spirals. Here, $\eta_i$ is defined as the local grain enhancement factor relative to the mean dust-to-gas ratio in the disc for the $i$th grain size. The local dust surface density for the $i$th grain size, $\Sigma_{{\rm d},i}$, will then be enhanced as,
\begin{equation}
    \Sigma_{{\rm d},i} = \langle\epsilon_i\rangle(\Sigma_0 + \eta_i\delta\Sigma),
\end{equation}
where $\langle\epsilon_i\rangle$ is the average dust-to-gas ratio for each grain size in the disc. Here we use the canonical value of $\langle\epsilon\rangle=0.01$ to represent the total dust-to-gas ratio over all grain sizes.

Particles with ${\rm St}_i \ll 1$ will be strongly coupled to the gas, experience minimal radial drift and will therefore have $\eta_i \approx 1$. The dust surface density will exactly trace the gas distribution in this case, with $\Sigma_{{\rm d},i} = \langle\epsilon_i\rangle(\Sigma_0 + \delta\Sigma)$. Large solids with ${\rm St}_i \gg 1$ will be entirely decoupled from the gas and will have constant surface density across the disc, with $\eta_i \approx 0$ and $\Sigma_{{\rm d},i} = \langle\epsilon_i\rangle\Sigma_0$. Note that we set a lower limit of $\eta_i = 0$ here. Intermediate sized grains with ${\rm St}_i \approx 1$ will generate peak enhancement factors of $\eta_i \approx 1 + d$, and therefore dust surface densities, $\Sigma_{{\rm d},i} = \langle\epsilon_i\rangle(\Sigma_0 + (1 + d)\delta\Sigma)$.

In Equation \ref{eq:dsigmamax}, regions coincident with the spiral peaks, where $m\phi = 0^{\circ}$, will experience maximum enhancement by a factor $\Sigma_0 + \eta_i\langle\delta\Sigma\rangle$, as $\delta\Sigma = \langle\delta\Sigma\rangle$ in these regions. Dust surface density in inter-arm regions, where $m\phi=180^{\circ}$, will equally be depleted by a factor $\Sigma_0 - \eta_i\langle\delta\Sigma\rangle$, as $\delta\Sigma = -\langle\delta\Sigma\rangle$ here.

To avoid $\Sigma_{{\rm d},i}$ becoming negative in inter-arm regions, we employ a correction factor,
\begin{equation}
  \Sigma_{{\rm d},i,{\rm corr}} = 
  \begin{cases}
      \eta_i\langle\delta\Sigma\rangle - \Sigma_0, & \text{if } \Sigma_0 + \eta_i\delta\Sigma < 0 \\
      0, & \text{otherwise}.
    \end{cases} 
\end{equation}
Thus our resultant dust surface density becomes,
\begin{equation}
    \Sigma_{{\rm d},i} = \frac{\langle\epsilon_i\rangle(\Sigma_0 + \eta_i\delta\Sigma + \Sigma_{{\rm d},i,{\rm corr}})\Sigma_0}{\Sigma_0 + \Sigma_{{\rm d},i,{\rm corr}}}.
\end{equation}

This ensures $\Sigma_{\rm d} > 0$ by increasing our dust distribution by a factor $\Sigma_{\rm d,i,corr}$ in cases where $\Sigma_0 + \eta_i\delta\Sigma < 0$. The denominator is a normalisation which ensures our mean dust surface density remains unchanged by $\Sigma_{\rm d,i,corr}$, thus ensuring mass conservation.

\subsection{Monte Carlo Radiative Transfer: \textsc{torus}}

Our disc is constructed within a mesh of grid cells, where initially we begin with a parent cell centred on the disc centre. We repeatedly subdivide parent cells into $2^D$ child cells based on some mass resolution criteria, where $D$ is the dimensions of our domain (3 dimensional here). If the mass in a cell exceeds $1\times 10^{-4}$\,M$_{\odot}$ then we further subdivide each cell into $2^D$ child cells such that child cells then become parent cells. This continues until the mass in each cell is less than or equal to our mass resolution criteria.

The dust temperatures are then calculated using the \textsc{torus} radiation transfer code \citep{harriesetal19}. Radiative equilibrium is calculated using the Monte Carlo technique originally described in \cite{lucy99}. Our discs are illuminated by a central star, whose radiation field is here represented by $10^9$ photon packets. These photon packets are emitted from the star isotropically and proceed to undergo a random walk through the grid, experiencing both absorption and scattering, until they escape the computational domain and the dust temperatures can be calculated assuming radiative equilibrium. Another cycle of $10^9$ photon packets are then emitted, now with these updated temperatures, until the dust temperatures are found to converge and continuum images can be produced.

\section{ALMA Simulations: \textsc{casa}}\label{sec:casa}

The output continuum images from \textsc{torus} are then used as inputs to the ALMA simulator in the Common Astronomy Software Application (\textsc{casa}) package (version 5.1) \citep{casa} to produce realistic synthetic ALMA images from our disc models. We use ALMA cycle 7 array configurations to produce these images, exploring various array sizes and resolutions in order to find optimal configurations for each observing frequency. 

We apply unsharp image masking \citep{malin77} to generate residual images from our synthetic observations by subtracting a smoothed radial profile of the image flux from itself. This technique highlights any non-axisymmetric features in our images, specifically spiral arms, by reducing the image flux range without reducing its dynamical range. We subtract a 2D Gaussian profile of FWHM closely matched to the beam size of our simulated images (we use 0.05"x0.05" here), and scaled with the peak image flux.

\begin{figure*}
    \includegraphics[width=\linewidth]{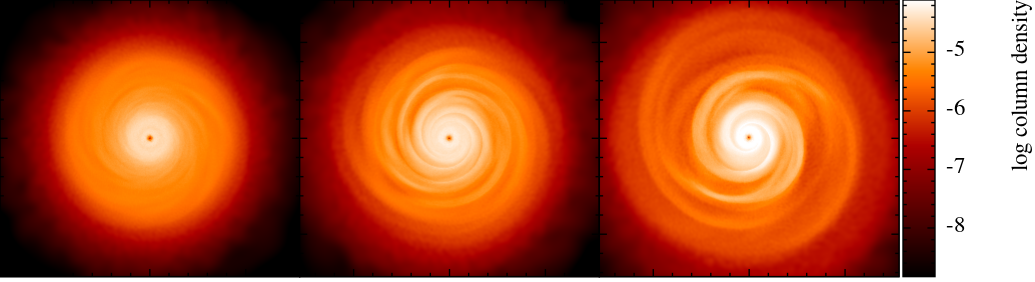}
    \caption{\label{fig:discinit} Surface density structure of self-gravitating SPH discs with $R_{\rm out}=100$\,AU after evolving for 5 outer orbital periods ($t=31420$\,yrs). Discs are constructed with 500,000 SPH gas particles and have mass ratios $q=0.2, 0.3, 0.4$ from left to right.}
\end{figure*}
\begin{figure}
    \centering
    \includegraphics[width=\linewidth]{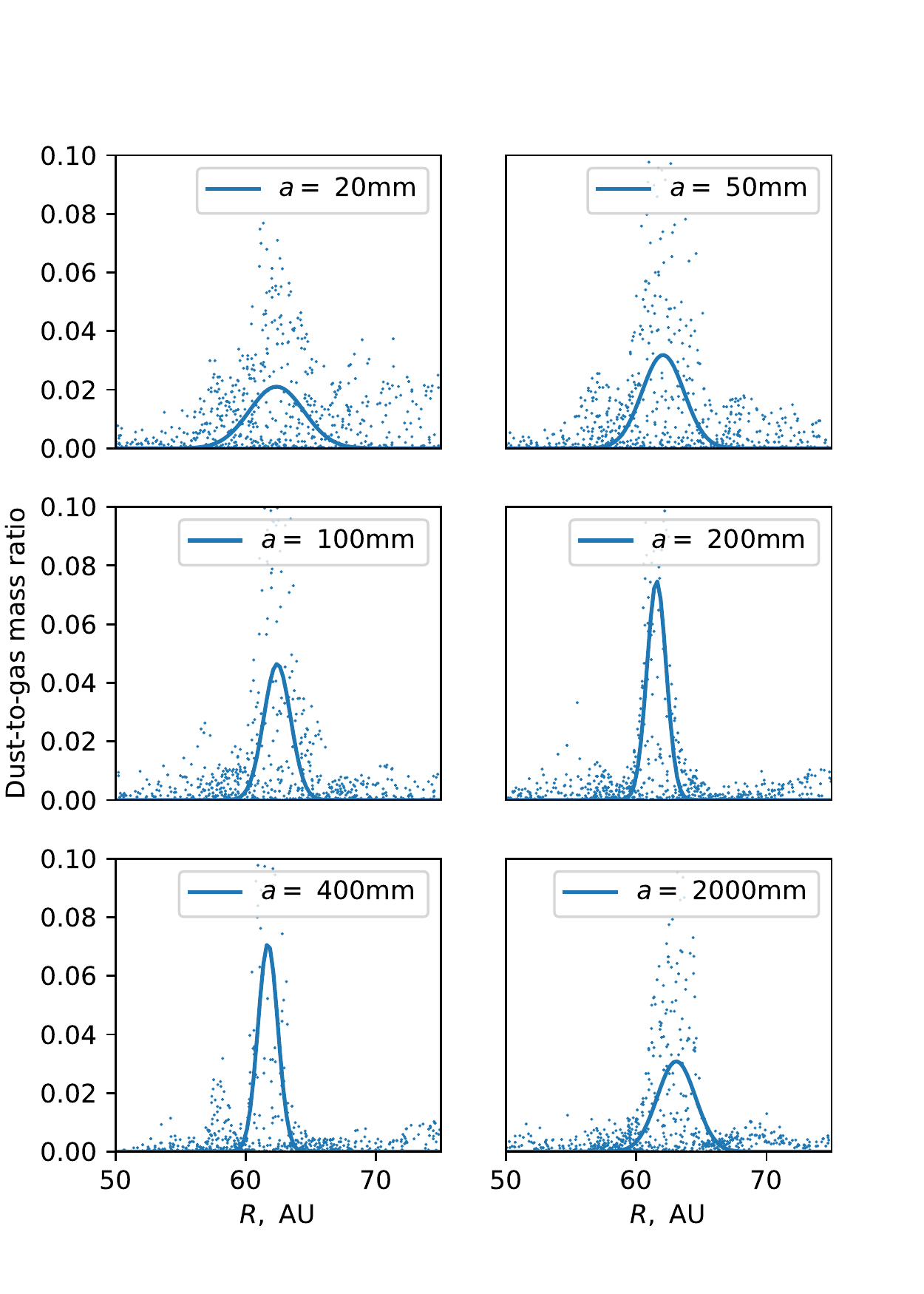}
    \caption{Gaussian fits to the dust-to-gas mass ratios taken from a radial slice of the $q=0.4$ disc, setup as described in Section \ref{sec:sphdust}. We plot how the dust-to-gas ratios vary for grain sizes $a=20$\,mm, $50$\,mm, $100$\,mm, $200$\,mm, $400$\,mm and $2000$\,mm. Grains sizes $a\approx200-500$\,mm become highly concentrated reaching peak dust-to-gas ratios $\epsilon\approx0.07$ here.}
    \label{fig:sphgrainconcdemo}
\end{figure}
\begin{figure}
    \includegraphics[width=\linewidth]{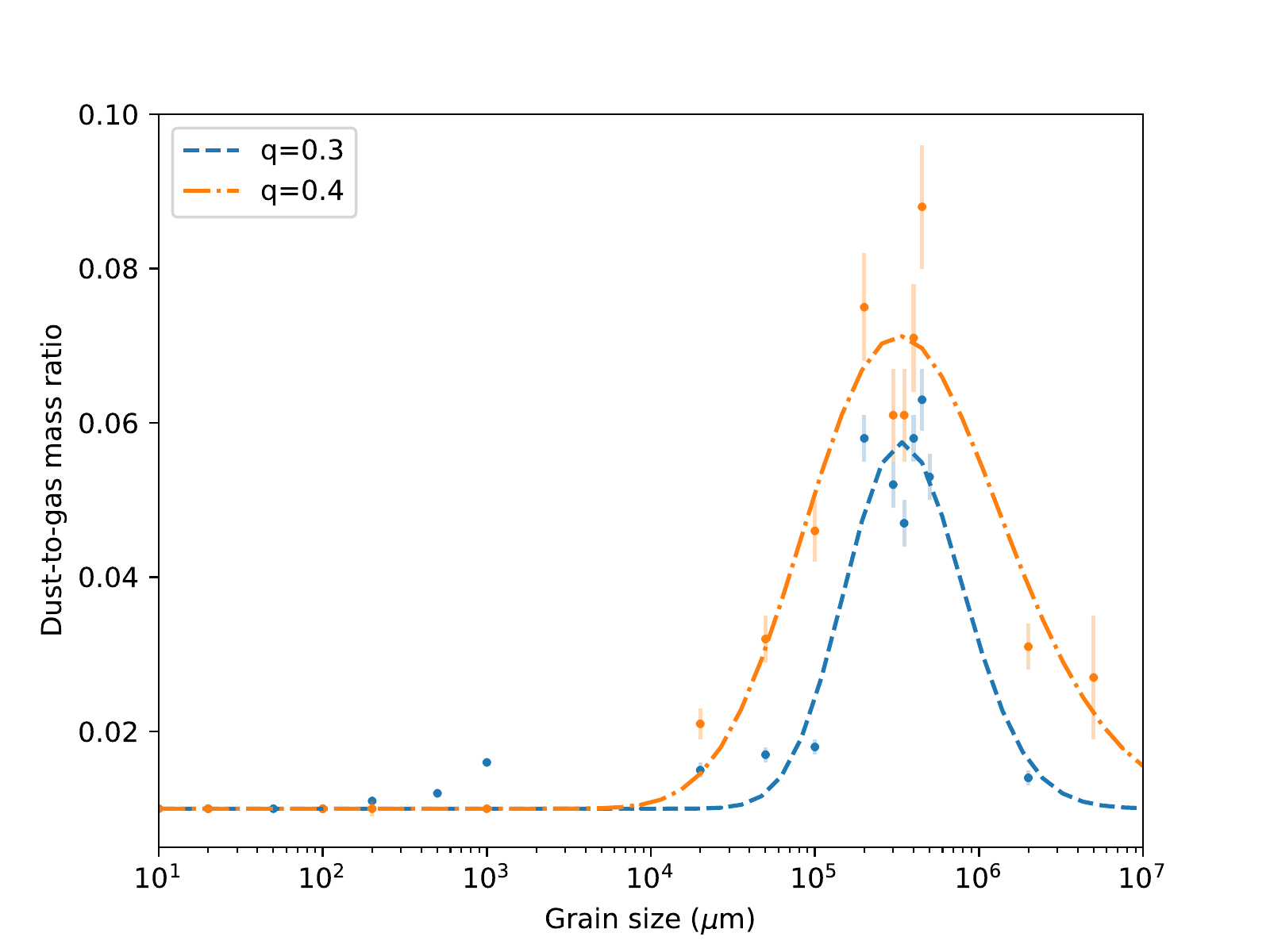}
    \caption{\label{fig:grainconc} Best-fit dust-to-gas ratios in SPH discs with mass ratios $q=0.3, 0.4$ and $R_{\rm out}=100$\,AU. Each disc consists of 500,000 gas particles, 125,000 dust particles and has been allowed to evolve for 6 outer orbital periods ($t=37700$yrs). We show the points with $1\sigma$ error bars obtained from their best-fit values. Log-normal curves are fitted to the data.}
\end{figure}

\section{SPH Models: Determining Peak Grain Enhancement}\label{sec:sphdust}

Our semi-analytic prescription of dust trapping in Equation \ref{eq:epsilon} requires that we determine the expected peak dust concentration factor, $d$, in disc spiral arms. To do this we employ the 3D SPH code \textsc{phantom} \citep{priceetal18} to numerically model the behaviour of dust particles in self-gravitating discs.

We set up three sets of discs with mass ratios $q=0.2$, $0.3$ and $0.4$ around a central star of mass $M_*=1$\,M$_{\odot}$. Each disc has initial inner and outer radii $R_{\rm in} = 1$\,AU and $R_{\rm out}=100$\,AU, and are set up with initial surface density profiles $\Sigma \propto R^{-1.5}$ and initial temperature profiles $T \propto R^{-0.5}$. We use artificial viscosity terms $\alpha_{\rm SPH}=0.1$ and $\beta_{\rm SPH}=0.2$. Cooling is modelled using the radiative transfer method introduced in \cite{stamatellosetal07}.

We use 500,000 SPH particles to represent the disc gas and we initially evolve the discs for 5 outer orbital periods with the gas only. We then inject a population of 125,000 dust SPH particles and allow the discs to evolve for a further orbital period. The final states of the gas-only discs are shown in Figure \ref{fig:discinit}. For each set of discs we run 20 separate simulations for 20 different grain sizes distributed log-normally between $0.1$\,${\rm \mu m}$ and $200$\,cm. To minimise computational expense, we neglect the self-gravity of these dust particles and treat them as test particles only.  

Dust-gas mixtures are modelled using two evolution models; the two-fluid method where the dust and gas are represented by two distinct particle populations coupled by a drag term \citep{laibeprice12a,laibeprice12b}, and the one-fluid method where the mixture is represented by gas particles only and the grain fraction is evolved along with the gas density for each particle  \citep{pricelaibe15}. The one-fluid method is implemented for smaller particle sizes at which the terminal velocity approximation is valid \citep[i.e. when the stopping time is shorter than the computational timestep, see][]{youdingoodman05}, thus it is not appropriate for modelling larger grains. We find an appropriate grain size boundary at which to switch between these two methods at $a\approx2$\,mm, therefore modelling all discs with $a \leq 2$\,mm using the one-fluid method, and discs with $a > 2$\,mm using the two-fluid method.

After evolving the dusty discs for a further orbital period, peak dust-to-gas ratios are determined by taking a radial slice of the disc, of azimuthal width $5^\circ$, and fitting a Gaussian distribution to the dust mass fraction at the spiral location. A demonstration of this is shown in Figure \ref{fig:sphgrainconcdemo}; we fit curves to a radial slice of the $q=0.4$ disc, where the spiral is located at $\approx60-70$\,AU. In Figure \ref{fig:grainconc} we fit log-normal curves to the best-fit dust-to-gas ratio peaks from the $q=0.3$ and $q=0.4$ discs. We exclude the $q=0.2$ disc from the remainder of this analysis as only weak spiral structure develops, therefore we observe only moderate grain enhancement.

Grain enhancement generally increases with increasing disc mass, primarily due to stronger spiral structure as we increase the disc-to-star mass ratio. This results in larger density gradients, greater radial drift velocities, and stronger concentration of grains. It is possible that grain concentration may continue to increase with increasing disc mass above $q=0.4$. However, for mass ratios $q \gtrsim 0.5$ discs become susceptible to fragmentation for the stellar mass considered here. This will act to disrupt any spiral arm structure thus limiting grain concentration. We therefore only model disc masses up to $q=0.4$.

Grains become most concentrated for sizes $a\approx200-500$\,mm, with peak dust-to-gas ratios $\epsilon\approx0.06$ and $\epsilon\approx0.07$ in the $q=0.3$ and $q=0.4$ discs respectively, giving values of  $d\approx5$ and  $d\approx6$ for equation \ref{eq:epsilon}. For the discs generated in Section \ref{sec:discparams}, with disc masses $q \lesssim 0.3$, we assume a maximum value of $d=5$ in our models.

\section{Grain Growth and the Fragmentation Threshold}\label{sec:grainsizes}

Appropriate grain size distributions for the equations in Section \ref{sec:discmodel} can be obtained using models of grain growth in protoplanetary discs. Grain growth proceeds through steady coagulation and accumulation during grain-grain collisions \citep{testi14}. The tendency for grains to stick together and grow during these collisions will depend on their collisional velocities. Particles with ${\rm St}<1$ (i.e. smaller grains) have smaller relative azimuthal velocities, hence when they collide they will likely coalesce in a so-called \textit{hit-and-stick} process \citep{chokshietal93,dominiktielens97}.

Larger particles will have higher relative azimuthal velocities, reaching a constant maximum value for ${\rm St}\geq 1$. \cite{kotheetal13} find a power-law mass dependence for the affinity of solids, $v_{\rm th}\propto m^{-3/4}$, with less massive solids having a greater threshold velocity for sticking. As particles grow, their impact velocities will increase accordingly and collisions will result in particles either bouncing off each other, compacting their densities in the process \citep{guttleratal10,zsometal10}, or shattering into several smaller fragments. These two growth barriers, known as the \textit{bouncing barrier} and the \textit{fragmentation threshold} respectively, may consequently limit the maximum size to which grains are able to grow through collisional accumulation, therefore limiting our value of $a_{\rm max}$.

The particle size at which the bouncing barrier is reached will depend on a number of factors such as particle porosity, density and material, and is therefore non-trivial to calculate analytically. Instead, we reason that the wealth of smaller, micron-sized solids dominating the dust-mass budget in discs \citep[see][]{williams11} requires regular replenishment through a cycle of growth and fragmentation, as otherwise these smaller grain sizes would quickly be depleted as they grow \citep{dullemondetal08, braueretal08, birnstielatal11}. This indicates that particles are able to grow to at least as large as the fragmentation threshold, and we therefore use this to define $a_{\rm max}$ in our models.

The fragmentation threshold velocity, $v_{\rm frag}$, is the maximum relative velocity that particles can withstand before collisions result in shattering. Relative azimuthal velocities scale with Stokes number, and for large Stokes' numbers particle's relative velocities will be dominated by turbulence. We can therefore calculate a maximum, threshold Stokes number for particles as \citep{birnstieletal10, birnstieletal12},
\begin{equation}
    {\rm St_{max}} \propto \frac{v_{\rm frag}^2}{\alpha c_{\rm s}^2},
\end{equation}
giving a maximum grain size of \citep{dipierro15},
\begin{equation}\label{eq:afrag}
    a_{\rm max} = \frac{4\langle\Sigma_{\rm g}\rangle}{3\pi\alpha\rho_{\rm s}}\frac{v_{\rm frag}^2}{\langle c_{\rm s}\rangle^2},
\end{equation}
where we use the azimuthally averaged gas surface density, $\langle\Sigma_{\rm g}\rangle$, and sound speed, $\langle c_{\rm s}\rangle$, as spiral features are short lived and grain growth timescales typically exceed these. We can estimate the viscous$-\alpha$ here by assuming that in a quasi-steady, self-gravitating disc dominated by turbulent motion, the viscous stress will saturate at a maximum value $\alpha=0.06$ \citep{riceetal05}, therefore defining the limiting maximum grain size.

\begin{figure}
    \includegraphics[width=\linewidth]{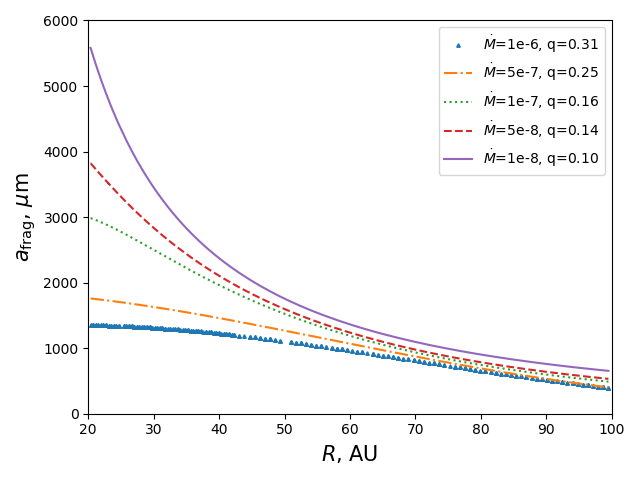}
    \caption{\label{fig:afrag10} Radial distribution of the fragmentation threshold from Equation \ref{eq:afrag} for mass accretion rates $\dot{\rm M} = 1 \times 10^{-6}$\,M${_{\odot}}$\,yr$^{-1}$, $5 \times 10^{-7}$\,M${_{\odot}}$\,yr$^{-1}$, $1 \times 10^{-7}$\,M${_{\odot}}$\,yr$^{-1}$, $5 \times 10^{-8}$\,M${_{\odot}}$\,yr$^{-1}$ and $1 \times 10^{-8}$\,M${_{\odot}}$\,yr$^{-1}$, and where $v_{\rm frag} = 10$\,ms$^{-1}$.}
\end{figure}

\begin{figure}
    \includegraphics[width=\linewidth]{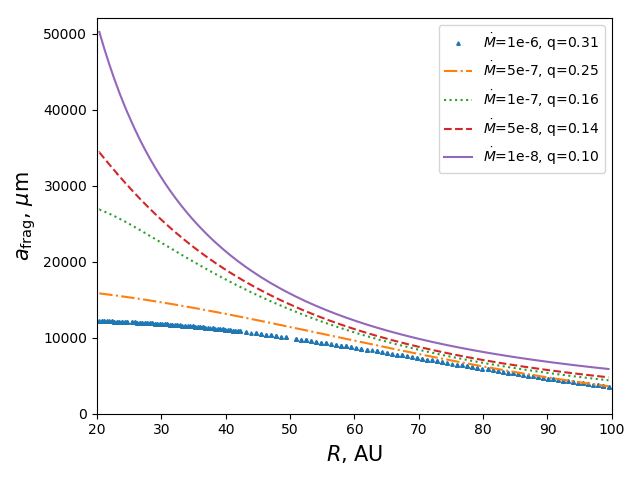}
    \caption{\label{fig:afrag30} Radial distribution of the fragmentation threshold from Equation \ref{eq:afrag} for mass accretion rates $\dot{\rm M} = 1 \times 10^{-6}$\,M${_{\odot}}$\,yr$^{-1}$, $5 \times 10^{-7}$\,M${_{\odot}}$\,yr$^{-1}$, $1 \times 10^{-7}$\,M${_{\odot}}$\,yr$^{-1}$, $5 \times 10^{-8}$\,M${_{\odot}}$\,yr$^{-1}$ and $1 \times 10^{-8}$\,M${_{\odot}}$\,yr$^{-1}$, and where $v_{\rm frag} = 30$\,ms$^{-1}$.}
\end{figure}

We use this to set our value of $a_{\rm max}$ in our disc models assuming two cases of $v_{\rm frag}=10$\,ms$^{-1}$ and $v_{\rm frag}=30$\,ms$^{-1}$. The mid-plane distributions of $a_{\rm frag}$ are plotted in Figures \ref{fig:afrag10} and \ref{fig:afrag30} for discs of outer radius, $R_{\rm out}=100$\,AU, and mass accretion rates ranging from $\dot{\rm M}=1 \times 10^{-8}$\,M${_{\odot}}$\,yr$^{-1}$ to $\dot{\rm M} = 1 \times 10^{-6}$\,M${_{\odot}}$\,yr$^{-1}$.


The fragmentation threshold decreases with increasing $\dot{\rm M}$ (i.e. with increasing disc mass), and becomes smallest in the outer disc where $a_{\rm frag}$ is comparable for all disc masses. In the most massive discs when $v_{\rm frag}=10$\,ms$^{-1}$, grains can only grow to $ \sim $\,mm sizes before collisions become destructive, with this maximum grain size in the inner disc decreasing by a factor of $\sim 5$ as we increase the disc mass from $q=0.1$ to $q=0.31$. For the higher threshold of $v_{\rm frag}=30$\,ms$^{-1}$ the value of $a_{\rm frag}$ increases by a factor $v_{\rm frag}^2$ for all disc masses (a factor 9), and grains can grow to $a_{\rm max}\sim$\,cm sizes here.

\section{Disc Models: Parameters}\label{sec:discparams}

With the additional information from Sections \ref{sec:sphdust} and \ref{sec:grainsizes}, it is now possible to use our models to efficiently predict for which disc parameters we expect self-gravitating disc substructure to be observable with ALMA. We setup discs as described in Section \ref{sec:discmodel} exploring a range of parameter space in disc masses, grain sizes and observing frequencies.

Our central star is modelled with $M_{*}=1$\,M$_{\odot}$, $R_{*}=2.325$\,R$_{\odot}$ and $T_{\rm eff}=4350$\,K. We assume a canonical dust-to-gas ratio of 0.01, and represent our grains as \cite{drainelee} silicates with size distribution,
\begin{equation}
    n(a) \propto a^{-q},
\end{equation}
distributed between minimum and maximum grain sizes $a_{\rm min}$ and $a_{\rm max}$, and assume $q=q_{\rm ism}=3.5$ \citep{mathisetal77}. We set here $a_{\rm min}=0.1$\,$\mu$m and vary the value of $a_{\rm max}$ to represent different stages of grain growth, using values $a_{\rm max}=10$\,${\rm \mu}$m (minimal grain growth), $1$\,mm, $10$\,cm, $100$\,cm, $a_{\rm frag, 10ms^{-1}}$ and $a_{\rm frag, 30ms^{-1}}$ (the grain fragmentation thresholds as described in Section \ref{sec:grainsizes}). We use 50 dust grain sizes distributed logarithmically between $0.1$\,$\mu$m and $2\times10^{6}$\,${\rm \mu}$m, and set the grain fraction for any grain size greater than $a_{\rm max}$ in each case to be zero.

We generate discs with 9 different mass accretion rates (equation \ref{eq:mdot}), where an increase in $\dot{M}$ roughly corresponds to an increase in disc mass. We use values of $\dot{M} = [1\times10^{-6}, 5\times10^{-7},2.81\times10^{-7},1.58\times10^{-7},1\times10^{-7},5\times10^{-8},2.81\times10^{-8},1.58\times10^{-8},1\times10^{-8}]$\,M$_{\odot}$yr$^{-1}$, which correspond to disc-to-star mass ratios, $q \approx 0.31,0.25,0.22,0.19,0.16,0.14,0.12,0.11$ and $0.10$ respectively. Using the relation between mass ratio and the number of spiral modes in equation \ref{eq:namrs}, and assuming a symmetrical response where we have an even number of modes, each of these discs are set up with $m=4$ and $m=8$ for the more massive and less massive cases respectively. A summary of these disc setups is laid out in Table \ref{tab:discsetup}.

\begin{table}
\centering
\begin{tabular}{|| c c c ||}
\hline\hline

\parbox{1cm}{\centering $\dot{M}$ \\ (M$_{\odot}$yr$^{-1}$)} & $M_{\rm disc}/M_*$ & $m$ \\
(1) & (2) & (3) \\

\hline
$1\times10^{-6}$ & 0.31 & 4 \\
$5\times10^{-7}$ & 0.25 & 4 \\
$2.81\times10^{-7}$ & 0.22 & 4 \\
$1.58\times10^{-7}$ & 0.19 & 4 \\
$1\times10^{-7}$ & 0.16 & 8 \\
$5\times10^{-8}$ & 0.14 & 8 \\
$2.81\times10^{-8}$ & 0.12 & 8 \\
$1.58\times10^{-8}$ & 0.11 & 8 \\
$1\times10^{-8}$ & 0.10 & 8 \\
 \hline\hline
\end{tabular}
\caption{(1) Mass accretion rates used for the discs setup in Section \ref{sec:discparams} and analysed in Section \ref{sec:taurusdiscresults}. (2) Calculated disc-to-star mass ratios. (3) Number of input spiral modes for each disc.}
\label{tab:discsetup}
\end{table}

\begin{figure*}
    \begin{tabular}{ccc}
    \centering
    \includegraphics[width=0.3\textwidth]{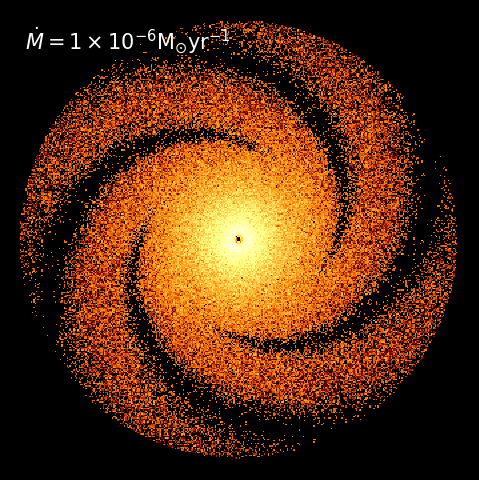} &
    \includegraphics[width=0.3\textwidth]{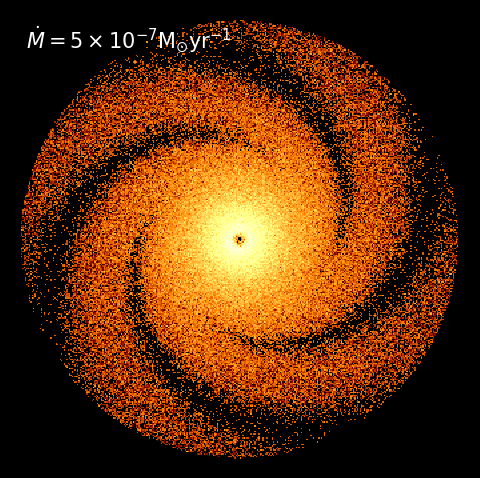} &
    \includegraphics[width=0.3\textwidth]{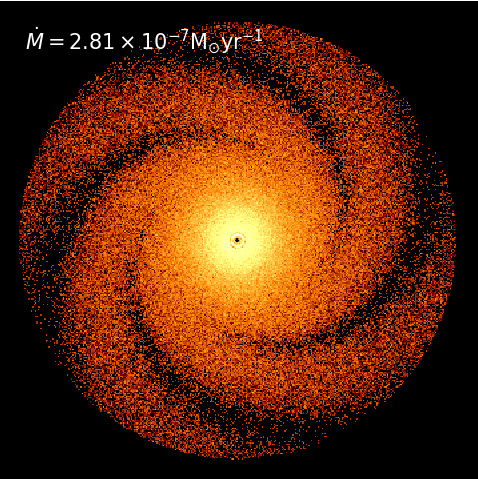} \\
    \centering
    \includegraphics[width=0.3\textwidth]{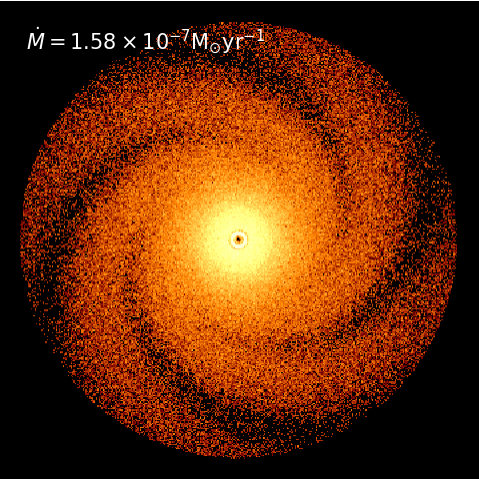} &
    \includegraphics[width=0.3\textwidth]{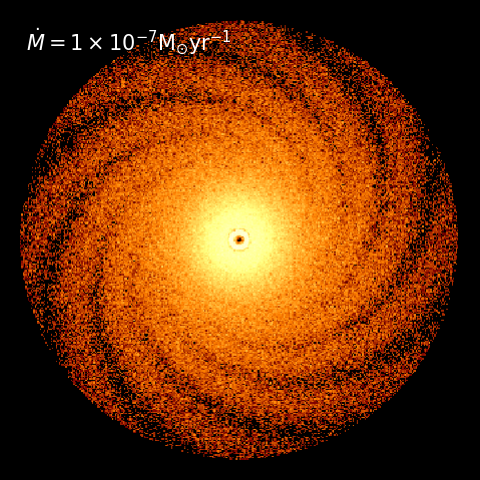} & \includegraphics[width=0.3\textwidth]{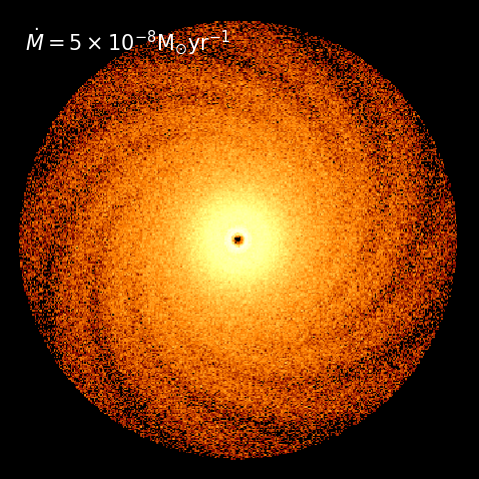} \\
    \centering
    \includegraphics[width=0.3\textwidth]{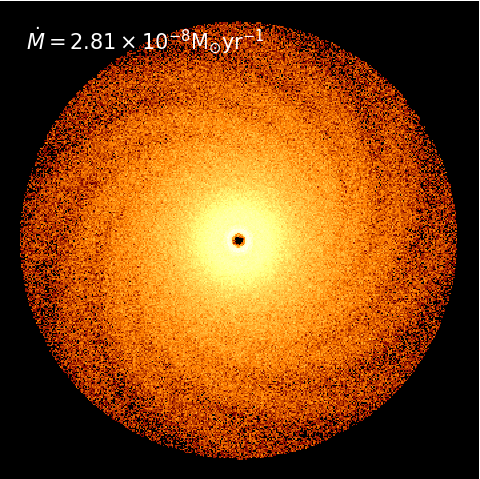} &
    \includegraphics[width=0.3\textwidth]{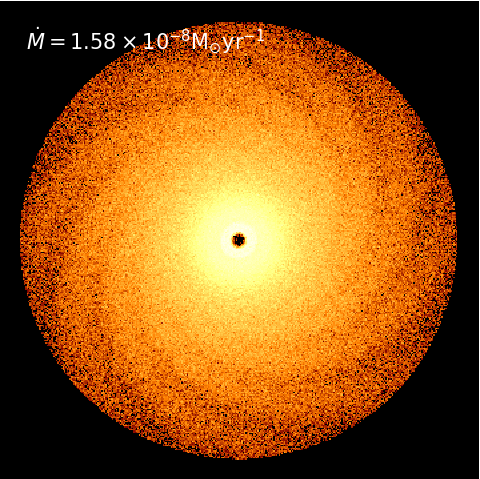} &
    \includegraphics[width=0.3\textwidth]{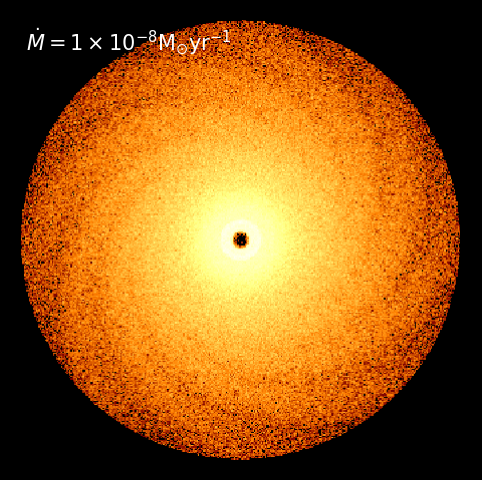} \\
    \end{tabular}
    \caption{\label{fig:ds9discs} \textsc{torus} disc continuum images at 230GHz ($\lambda=1.3$\,mm). Discs are set up with $R_{\rm out}=100$\,AU, grain size distributions $n(a) \propto a^{-3.5}$ with $a_{\rm min}=0.1$\,$\mu$m and $a_{\rm max}=1$\,mm, and mass accretion rates (from left to right) Top: $\dot{\rm M}$ = $1 \times 10^{-6}$\,M${_{\odot}}$\,yr$^{-1}$, $5 \times 10^{-7}$\,M${_{\odot}}$\,yr$^{-1}$, $2.81 \times 10^{-7}$\,M${_{\odot}}$\,yr$^{-1}$. Middle: $1.58 \times 10^{-7}$\,M${_{\odot}}$\,yr$^{-1}$, $1 \times 10^{-7}$\,M${_{\odot}}$\,yr$^{-1}$, $5 \times 10^{-8}$\,M${_{\odot}}$\,yr$^{-1}$. Bottom: $2.81 \times 10^{-8}$\,M${_{\odot}}$\,yr$^{-1}$, $1.58 \times 10^{-8}$\,M${_{\odot}}$\,yr$^{-1}$, $1 \times 10^{-8}$\,M${_{\odot}}$\,yr$^{-1}$.}
\end{figure*}

Continuum images of these discs are generated for observing frequencies $115$\,GHz (${\rm \lambda=2.6}$\,mm), $230$\,GHz (${\rm \lambda=1.3}$\,mm) and $690$\,GHz (${\rm \lambda=0.4}$\,mm), corresponding to ALMA observing bands 3, 6 and 9 respectively. We consider discs at a distance of 140pc, comparable to those in the Taurus star forming region. Example \textsc{torus} output images produced in this way are shown in Figure \ref{fig:ds9discs} for discs with $a_{\rm max}=1$\,mm and accretion rates from Table \ref{tab:discsetup}.

\begin{figure*}
    \raisebox{0.15\height}{\includegraphics[width=.3025\linewidth]{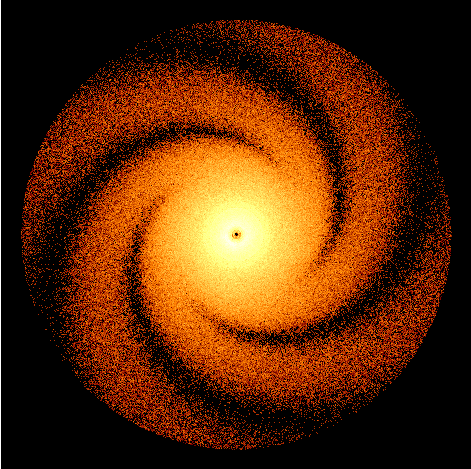}}
    \includegraphics[width=.33\linewidth]{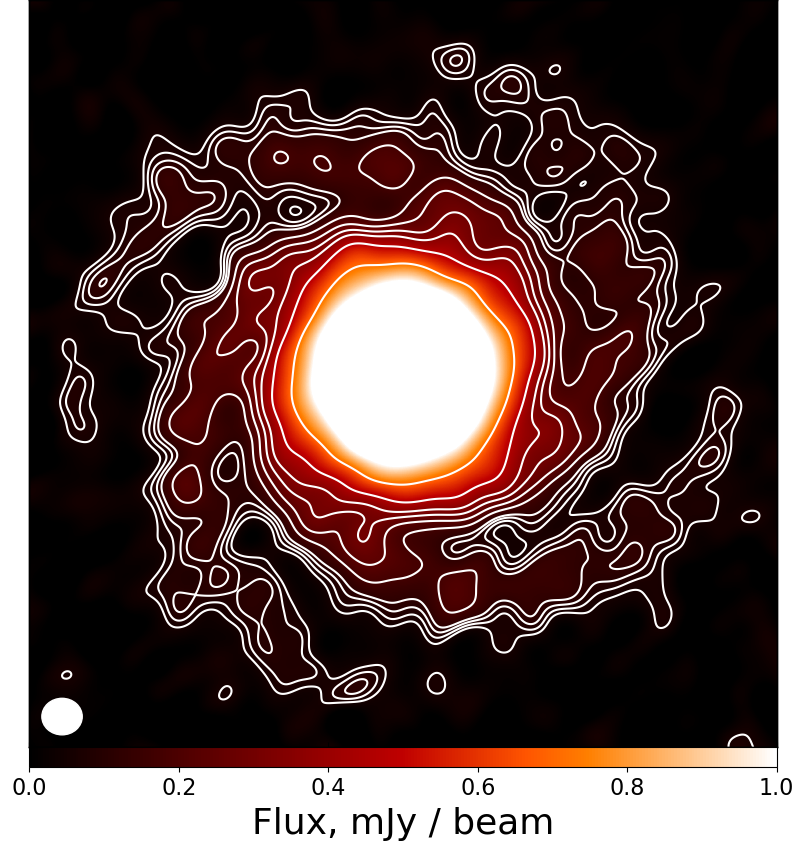}
    \includegraphics[width=.33\linewidth]{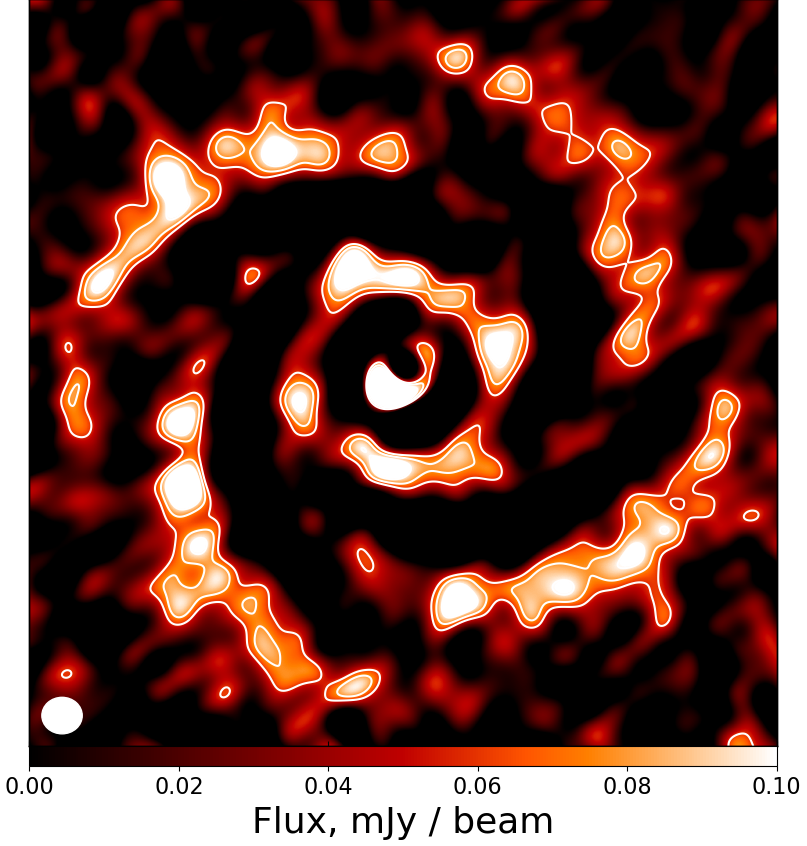}
    \caption{\label{fig:casademo} Demonstration of the process generating unsharp masked disc images from \textsc{torus} radiation transfer continuum profiles. Left: Output continuum disc image from \textsc{torus}. Middle: Synthetic ALMA observation using \textsc{casa}. Right: Unsharp masked residual image. Discs have properties $\dot{M}=5\times10^{-7}$\,M$_{\odot}$yr$^{-1}$, $R_{\rm out}=100$\,AU, $a_{\rm max}=1$\,mm and are observed at $f_{\rm obs}=115$\,GHz ($\lambda=2.6$\,mm) with observation exposure time, array configuration and PWV level laid out in Table \ref{tab:casa}.}

\end{figure*}

We then use these continuum disc images as inputs to the \textsc{casa} tasks \textsc{simobserve} and \textsc{simanalyze} and generate synthetic ALMA observations. Observing times, antenna configurations and PWV values used as inputs to \textsc{casa} are laid out in Table \ref{tab:casa}. Unsharp image masking is applied to these synthetic observations in order to highlight any non-axisymmetric disc features present, as described in Section \ref{sec:casa}. We demonstrate the process of generating synthetic ALMA observations and then unsharp masked residual images from \textsc{torus} continuum profiles in Figure \ref{fig:casademo}.

\begin{table}
\centering
\begin{tabular}{|| c || c c c ||}
\hline
$f_{\rm obs}$ & $t_{\rm obs}$ & Antenna Config & PWV Level \\
(1) & (2) & (3) & (4) \\
\hline\hline
\\
115\,GHz & 1800\,s & alma.cycle7.8 & 5.186\,mm \\
 \\
230\,GHz & 1800\,s & alma.cycle7.8 & 1.796\,mm \\  
 \\
690\,GHz & 1800\,s & alma.cycle7.6 & 0.472\,mm \\ 
 \\
 \hline
\end{tabular}
\caption{Input parameters used here for generating synthetic images with \textsc{casa}. (1) ALMA observing frequency. (2) Simulated observing time. (3) ALMA antenna configuration used. (4) Precipitable Water Vapour (PWV) level.}
\label{tab:casa}
\end{table}

\section{Disc Models: Results}\label{sec:taurusdiscresults}

Our focus here is to analyse the parameter space in which self-gravitating disc structure may be observable with ALMA. We present our results in this section considering the effects of varying disc mass, grain size distribution and observing frequency on our ability to distinguish spiral structure in our disc model. Galleries of unsharp masked synthetic disc images where we explore this parameter space can be found in Appendix \ref{appendix:gallery}.

\subsection{Analysing the impact of grain enhancement}

We begin this section by first demonstrating the impact of grain enhacement on observability. We showed in Section \ref{sec:sphdust} that dust trapping of $\sim$cm sized grains significantly enhances dust-to-gas ratios in spiral arm regions, therefore equally acting to remove dust from interarm regions. Spiral structure consequently becomes sharper and more distinct, producing higher flux ratios between arm and interarm regions due to enhanced and depleted emission at these locations respectively.

\begin{figure*}
    \centering
    \includegraphics[height=55mm]{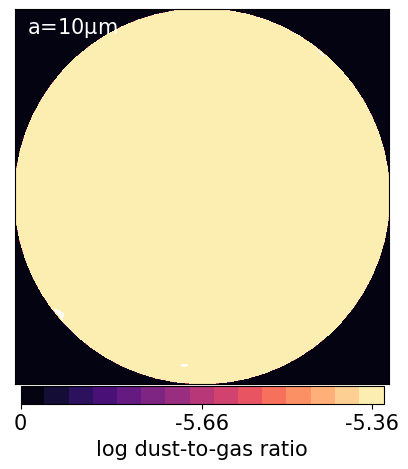}
    \includegraphics[height=55mm]{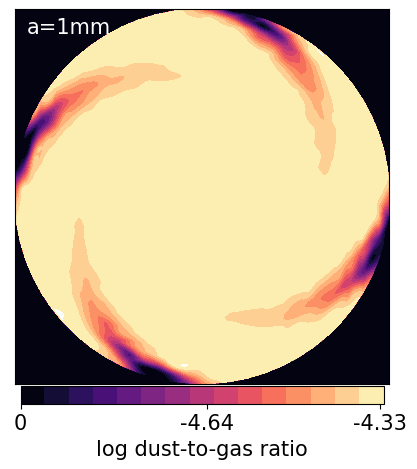}
    \includegraphics[height=55mm]{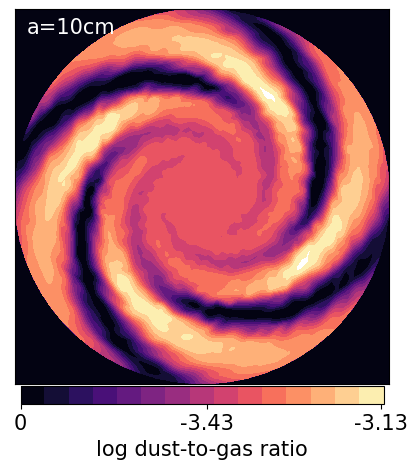}
    \caption{\label{fig:grainsizedist} Plotted are the dust-to-gas ratios for individual grain species of different sizes in a disc with $\dot{M}=1\times10^{-6}$\,M$_{\odot}$yr$^{-1}$, $R_{\rm out}=100$\,AU and grain size distribution $n(a) \propto a^{-3.5}$ with $a_{\rm min}=0.1\mu$m and $a_{\rm max}=100$\,cm. We plot the distributions for grains of sizes $a=10\mu$m (Left), $a=1$\,mm (Middle) and $a=10$\,cm (Right). We demonstrate the impact of our grain enhancement prescription outlined in Section \ref{sec:dustconc} as $\sim$cm sized grains become highly concentrated in the disc spiral arms. Note that the colourbars are scaled to the maximum dust-to-gas mass ratio in each respective grain size bin, $\epsilon_{i,\rm max}$.}
\end{figure*}

We illustrate our grain enhancement prescription in Figure \ref{fig:grainsizedist} by plotting how dust-to-gas ratio varies across our disc model for grains with sizes of $a=10$\,${\rm \mu}$m, $1$\,mm and $10$\,cm, in a disc with $\dot{M}=1\times10^{-6}$\,M$_{\odot}$yr$^{-1}$, $R_{\rm out}=100$\,AU and grain size distribution $n(a) \propto a^{-3.5}$ with $a_{\rm min}=0.1\mu$m and $a_{\rm max}=100$\,cm. Grains of $a=10{\rm \mu}$m with ${\rm St} \ll 1$ exactly trace the gas distribution and display an entirely uniform dust-to-gas ratio across the disc. The Stokes number, and therefore also grain concentration factor, $\eta$, scales with grain size up to ${\rm St}=1$. As we consider larger grain sizes up to $a=10$\,cm, grains become enhanced in the spirals arms and clear non-axisymmetric dust-to-gas ratios start to emerge.

\begin{figure*}
    \includegraphics[height=43mm]{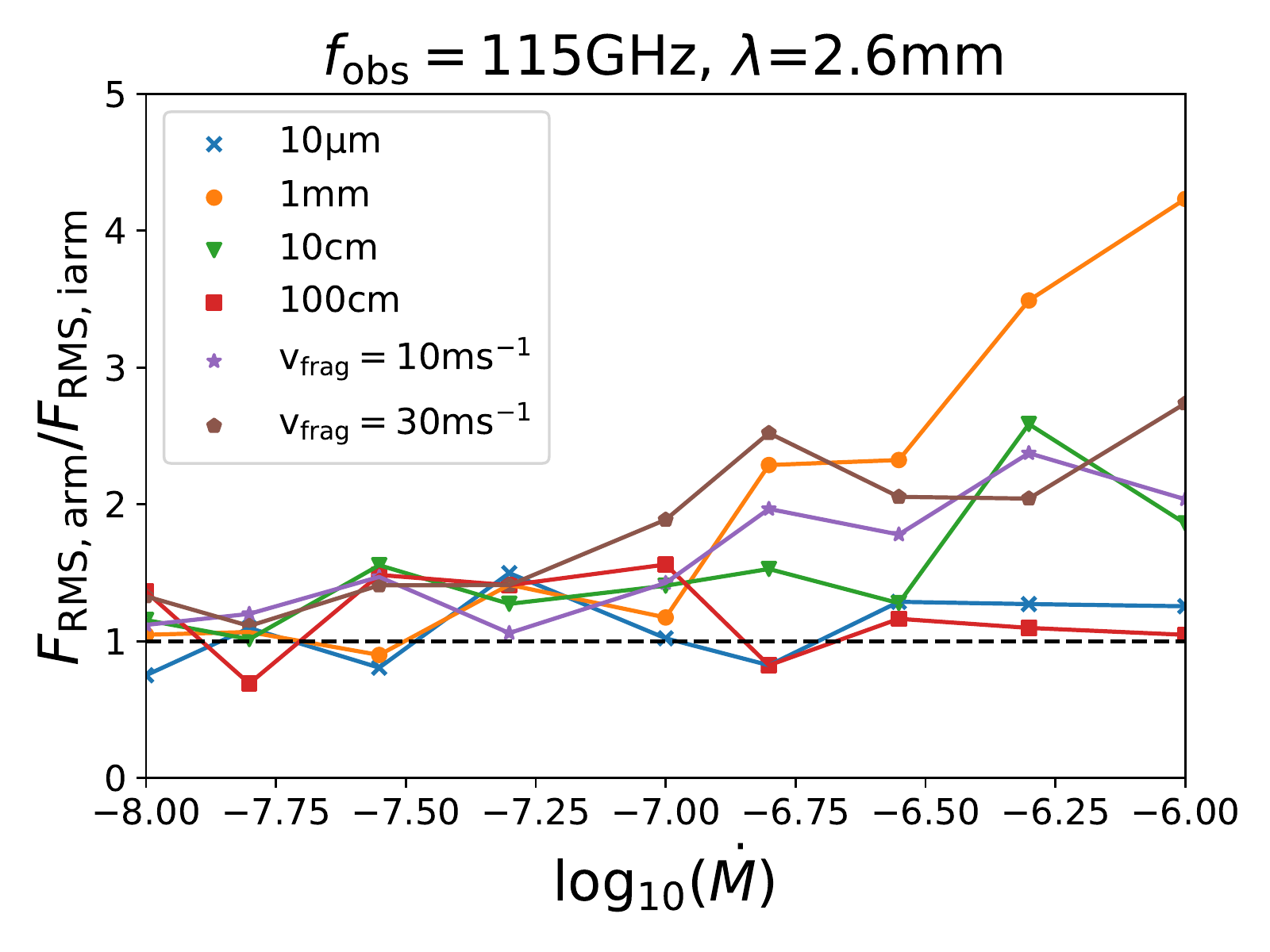}
    \includegraphics[height=43mm]{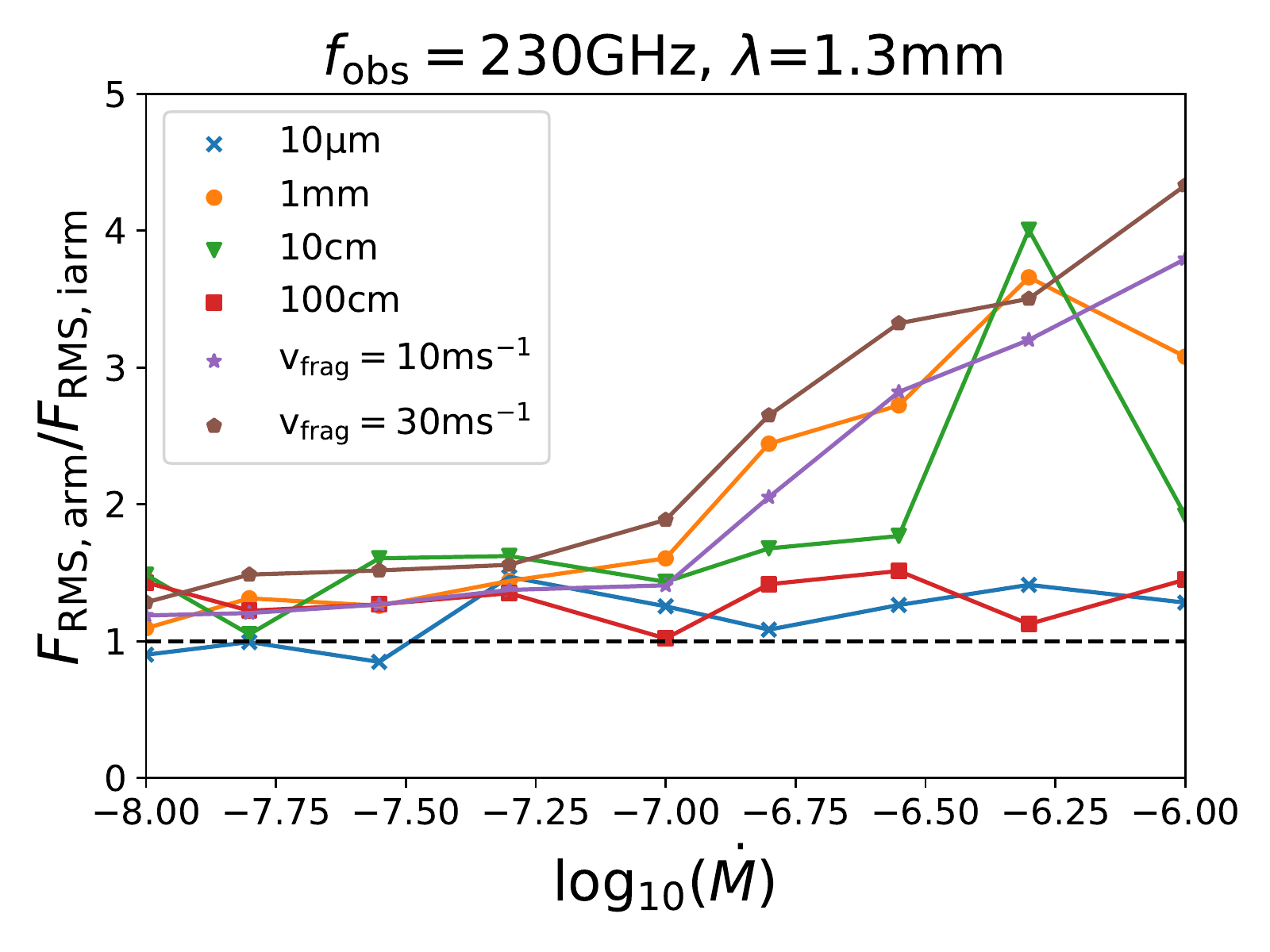}
    \includegraphics[height=43mm]{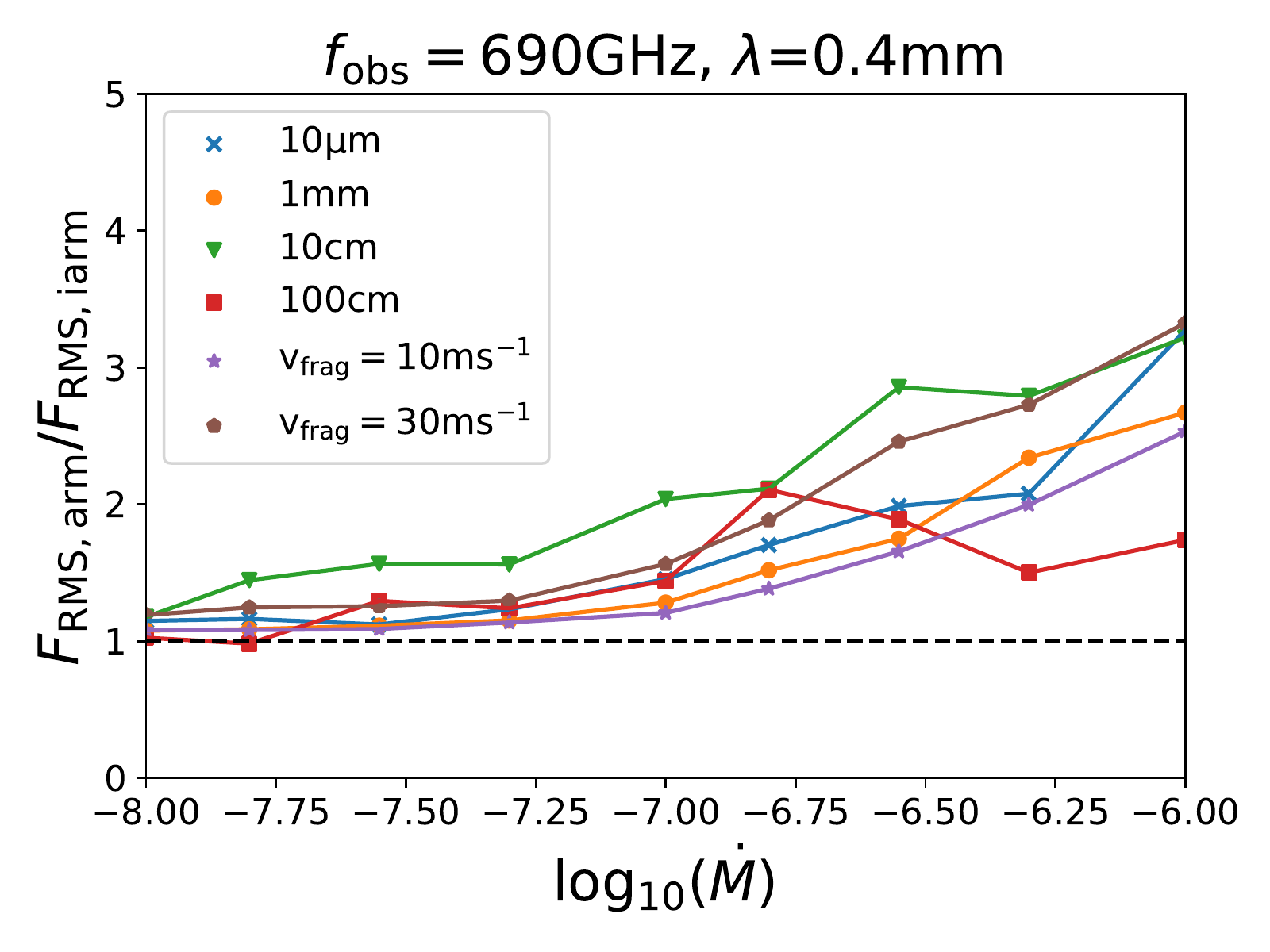}
    \caption{\label{fig:fluxvs.q} Ratios of RMS fluxes in spiral arm regions to RMS fluxes in interarm regions ($F_{\rm RMS,arm} /F_{\rm RMS,iarm}$) plotted against mass accretion rate, ${\rm log_{10}(}\dot{M})$, for the discs modelled in Section \ref{sec:discparams} and presented in Appendix \ref{appendix:gallery}. These plots are generated using the synthetic ALMA observations prior to performing unsharp image masking.}
\end{figure*}

\begin{figure*}
    \includegraphics[height=43mm]{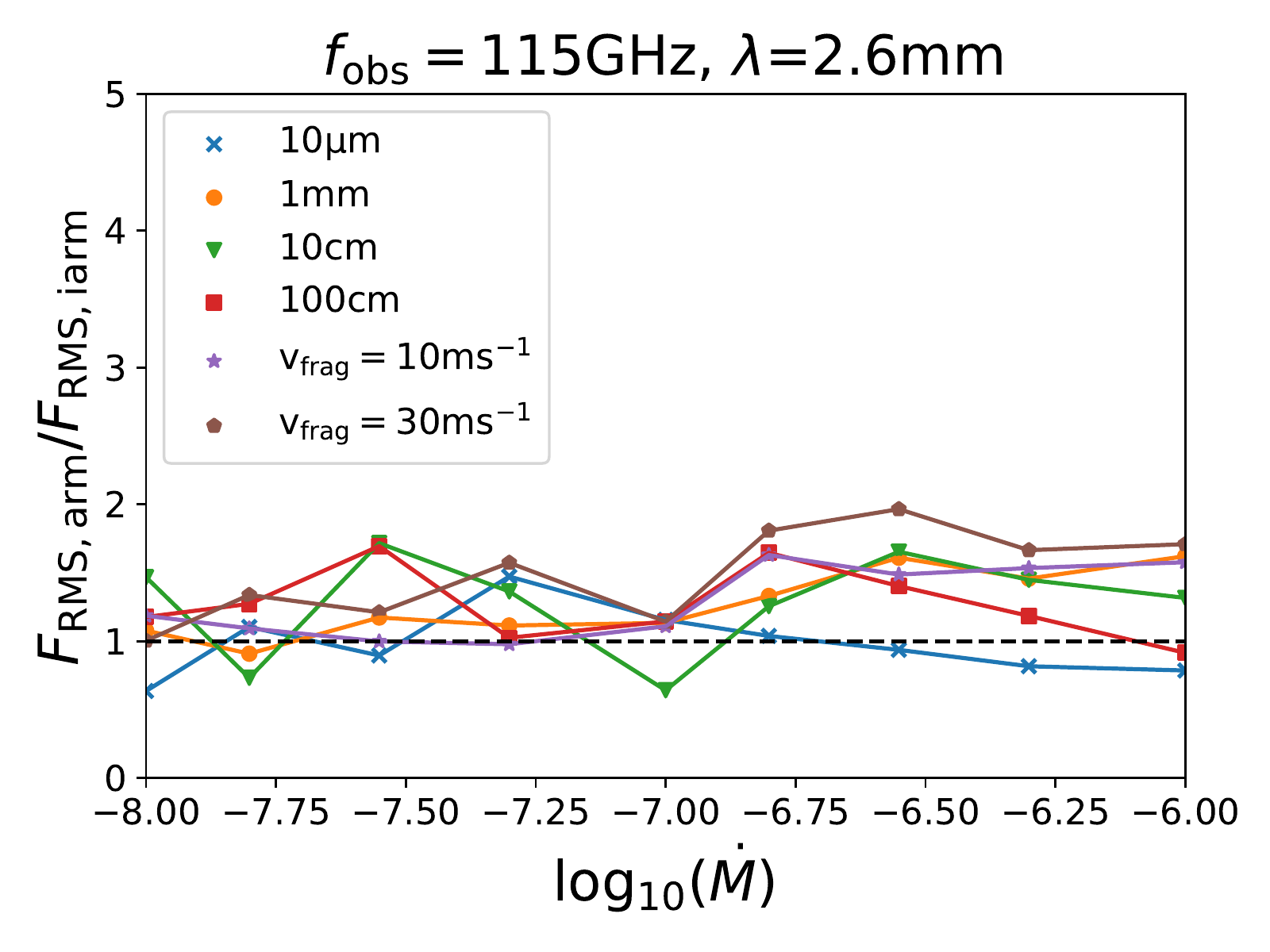}
    \includegraphics[height=43mm]{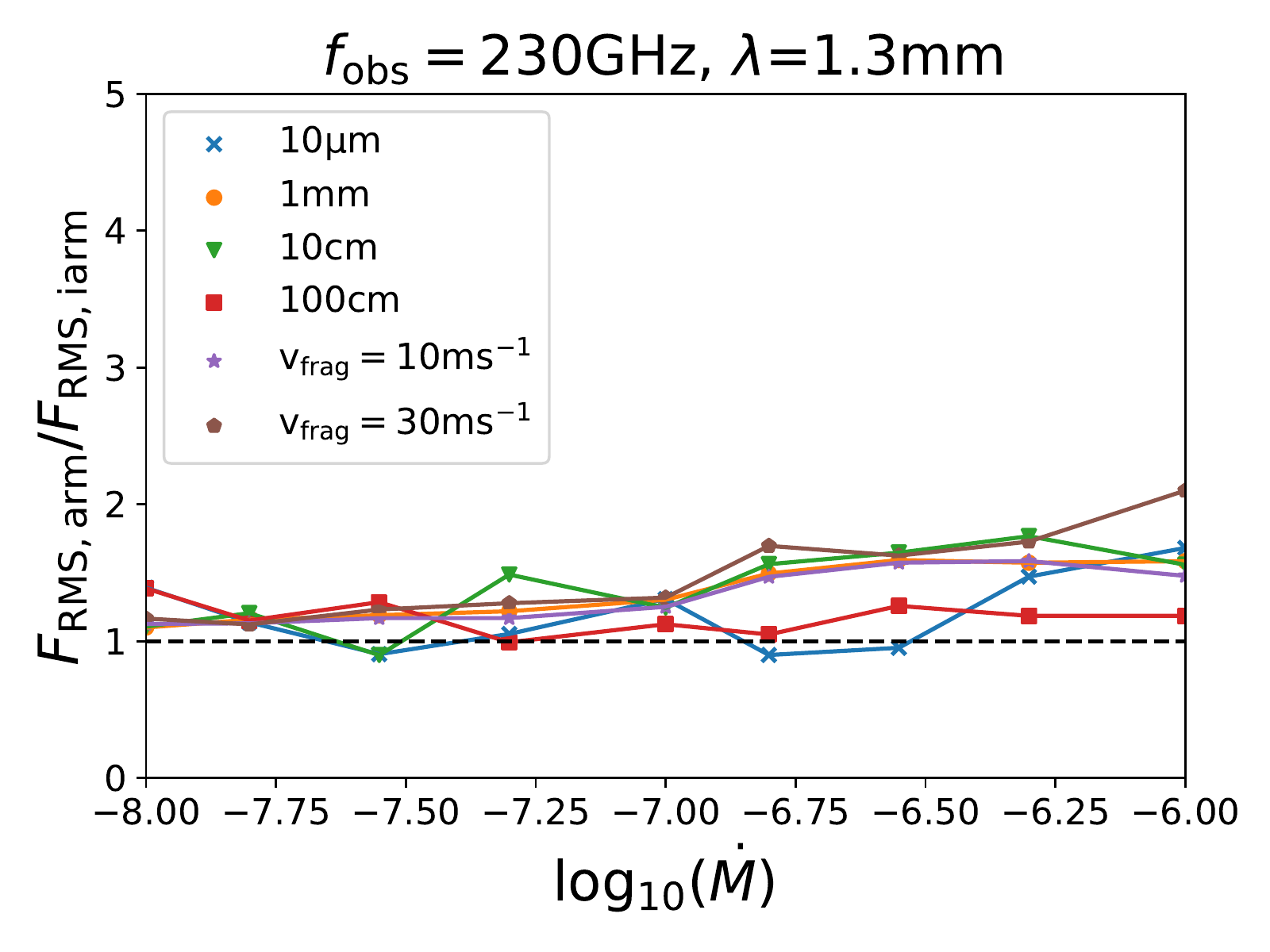}
    \includegraphics[height=43mm]{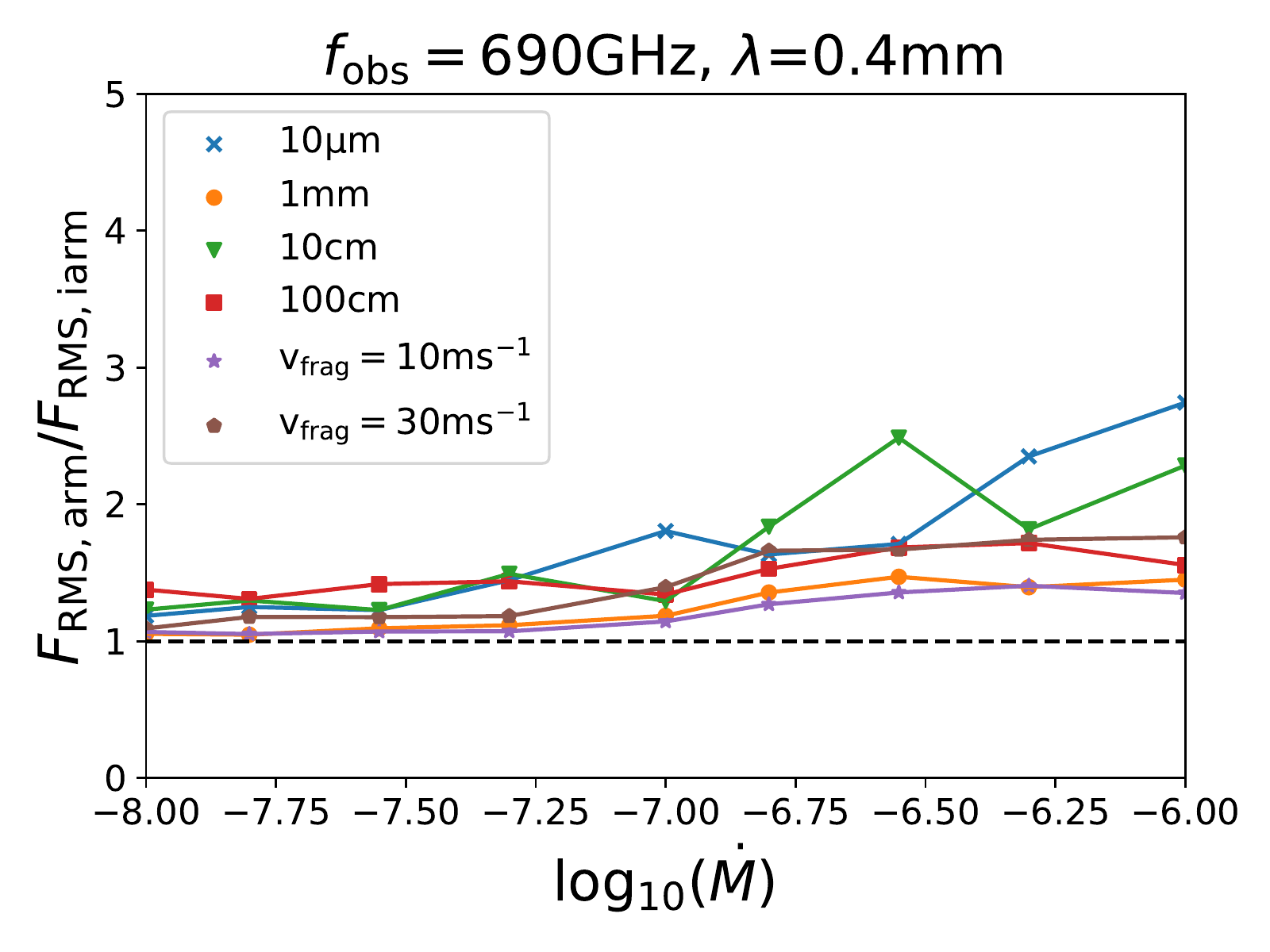}
    \caption{\label{fig:fluxvs.q-noconc} Ratios of RMS fluxes in spiral arm regions to RMS fluxes in interarm regions ($F_{\rm RMS,arm} /F_{\rm RMS,iarm}$) plotted against mass accretion rate, ${\rm log_{10}(}\dot{M})$, for the discs modelled in Section \ref{sec:discparams}. Here we do not account for grain enhancement in spiral arm regions, therefore reducing the prominence of spiral structure in discs compared to their counterparts in Figure \ref{fig:fluxvs.q}. These plots are generated using the synthetic ALMA observations prior to performing unsharp image masking.}
\end{figure*}

\begin{figure}
    \centering
    \includegraphics[height=35mm]{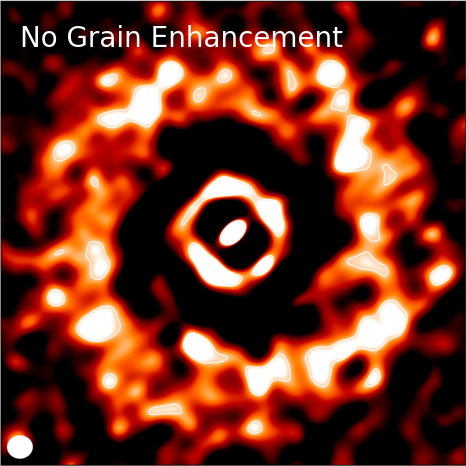}
    \includegraphics[height=35mm]{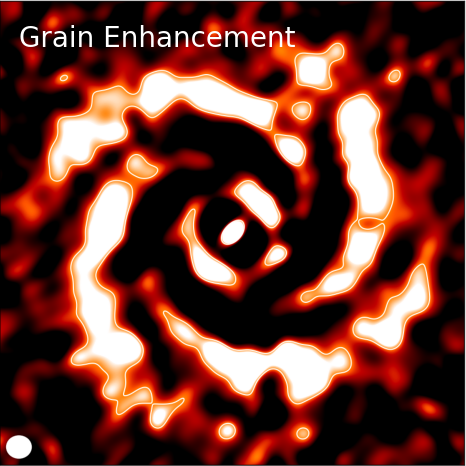}
    \includegraphics[height=35mm]{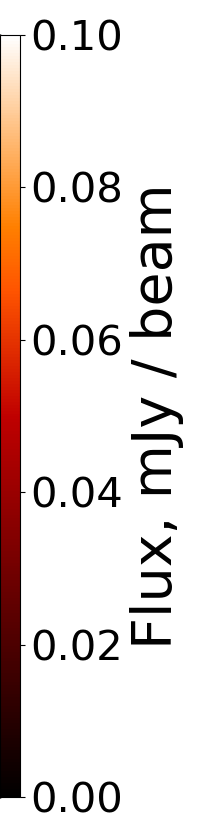}
    \includegraphics[height=35mm]{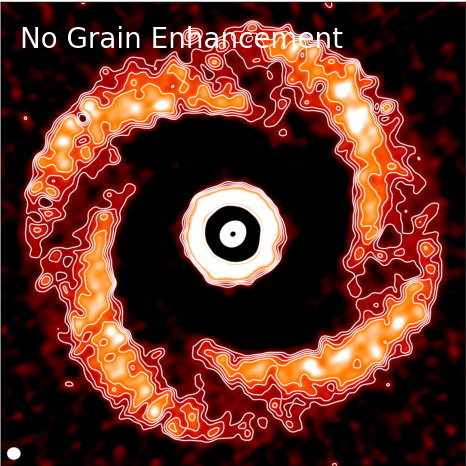}
    \includegraphics[height=35mm]{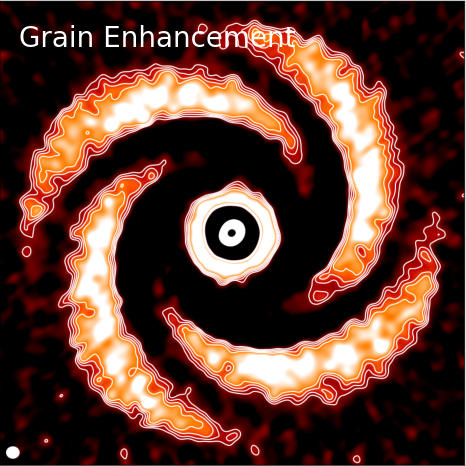}
    \includegraphics[height=35mm]{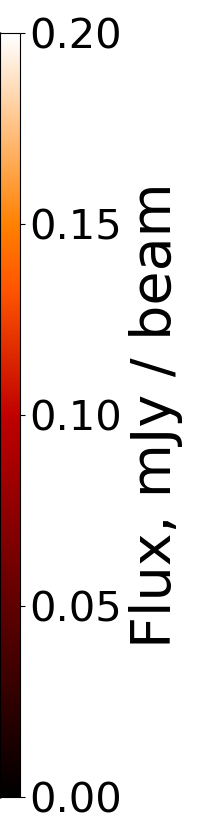}
    \includegraphics[height=35mm]{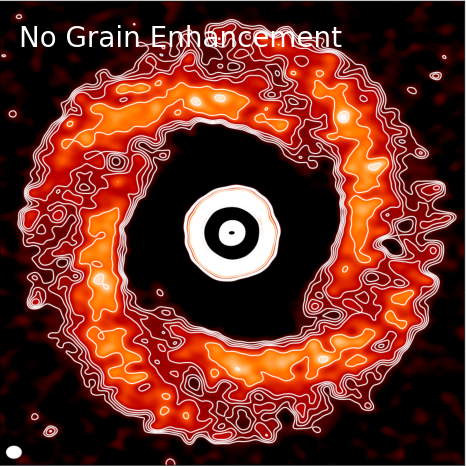}
    \includegraphics[height=35mm]{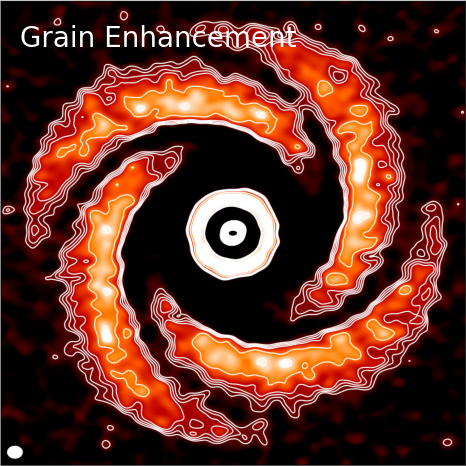}
    \includegraphics[height=35mm]{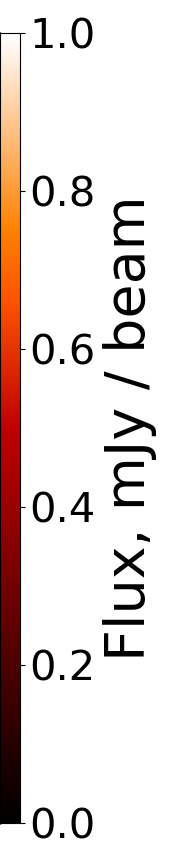}
    \caption{\label{fig:concvs.noconc} Unsharp masked disc images for $\dot{\rm M}=1\times10^{-6}$\,M$_{\odot}$\,yr$^{-1}$ and $a_{\rm max}=v_{\rm frag,30ms^{-1}}$, observed at frequencies Top: $115$\,GHz ($\lambda=2.6$\,mm), Middle: $230$\,GHz ($\lambda=1.3$\,mm) and Bottom: $690$\,GHz ($\lambda=0.4$\,mm). We compare like for like disc models with our prescription for grain enhancement included (right column) and not included (left column) in the disc models.}
\end{figure}

It is useful here to quantify observability of spiral structure in terms of the ratio of the RMS fluxes in the disc arm and interarm regions (i.e.
$F_{\rm RMS,arm} /F_{\rm RMS,iarm}$). Arm and interarm regions in our resultant disc images can be located using equation \ref{eq:spiraltheta}, and we calculate the RMS fluxes between radii $70-100$\,AU where we find spiral structure to be most prominent. In Figure \ref{fig:fluxvs.q} we plot how these flux ratios vary with mass accretion rate, and show comparison plots for models that do not include dust grain enhancement in Figure \ref{fig:fluxvs.q-noconc}. Flux ratios are calculated using the synthetic ALMA observations prior to unsharp masking. Example like-for-like unsharp masked disc images are also included for reference in Figure \ref{fig:concvs.noconc}.

For the same disc parameters we calculate considerably higher flux ratios when including dust trapping in our model, most notably when the dust mass budget is dominated by millimetre/centimetre grains (i.e. when $a_{\rm max}=$\,mm$-$cm sizes). Previously blurred arm and interarm regions become distinct as millimetre emission is concentrated in the spiral peaks. The key implication here is that with grain enhancement generating stronger spiral structure for the same mass discs, we should expect to detect self-gravitating disc structure for lower disc masses than previously predicted, if sufficient grain growth has occurred. In discs with no grain growth, or in which grains have grown well beyond centimetre sizes, the lack of dust mass in millimetre/centimetre aggregates becomes detrimental to the observability of disc substructure.

Given the short potential lifetime of a disc's self-gravitating phase its important to note how fast grains can actually grow, and thus what likely maximum grain size would be present in young, self-gravitating discs. Models of grain growth predict that millimetre and centimetre-sized grains form rapidly on timescales $\lesssim10^5$\,yrs \citep{dullemondetal05,laibetal08}. It therefore seems reasonable  to expect grains to have grown to at least as large as the fragmentation threshold before the end of a disc's self-gravitating phase, and that enhanced emission in spiral regions from these larger grains may be significant.

\begin{figure*}
    \centering
    \includegraphics[height=55mm]{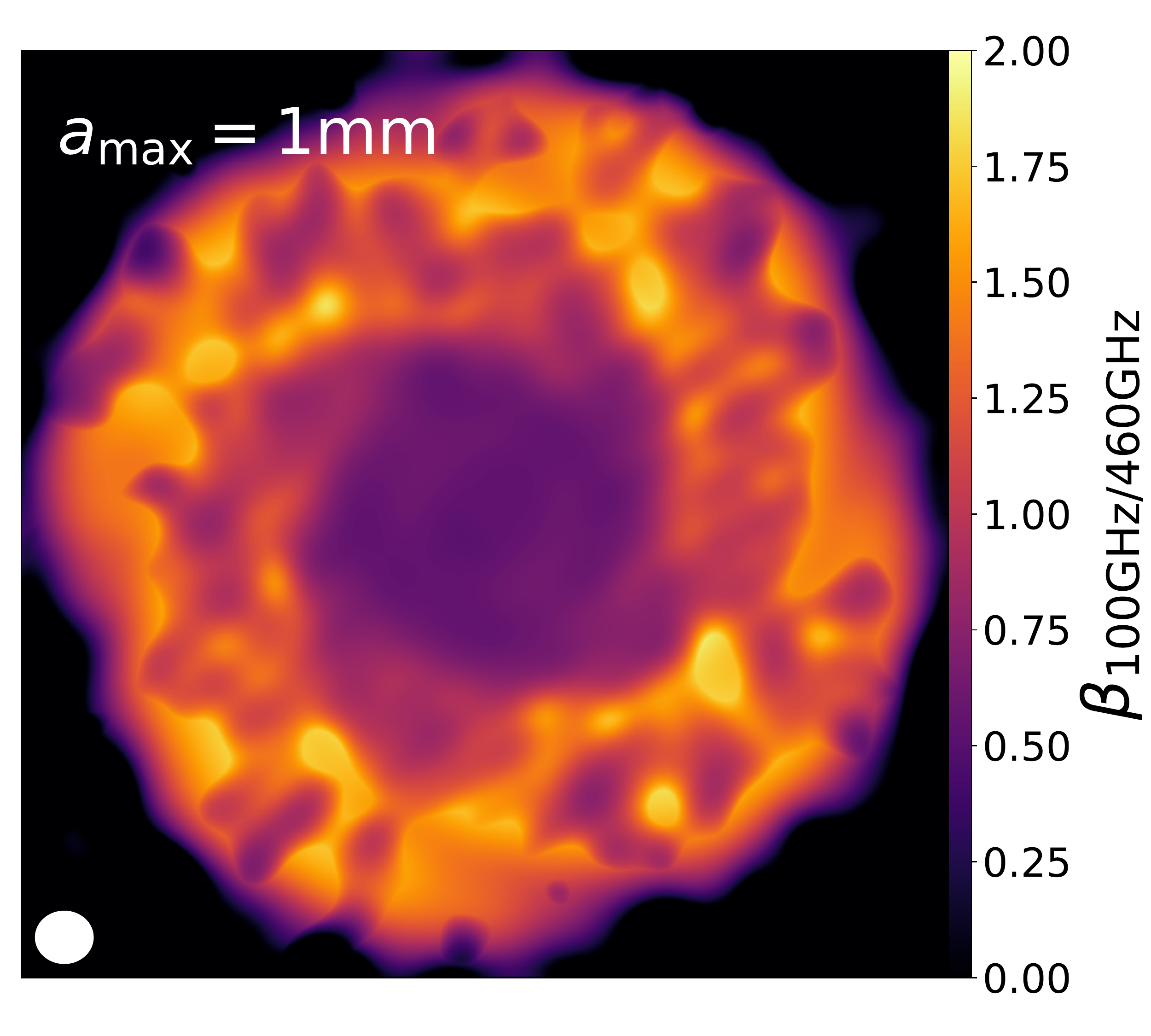}
    \includegraphics[height=55mm]{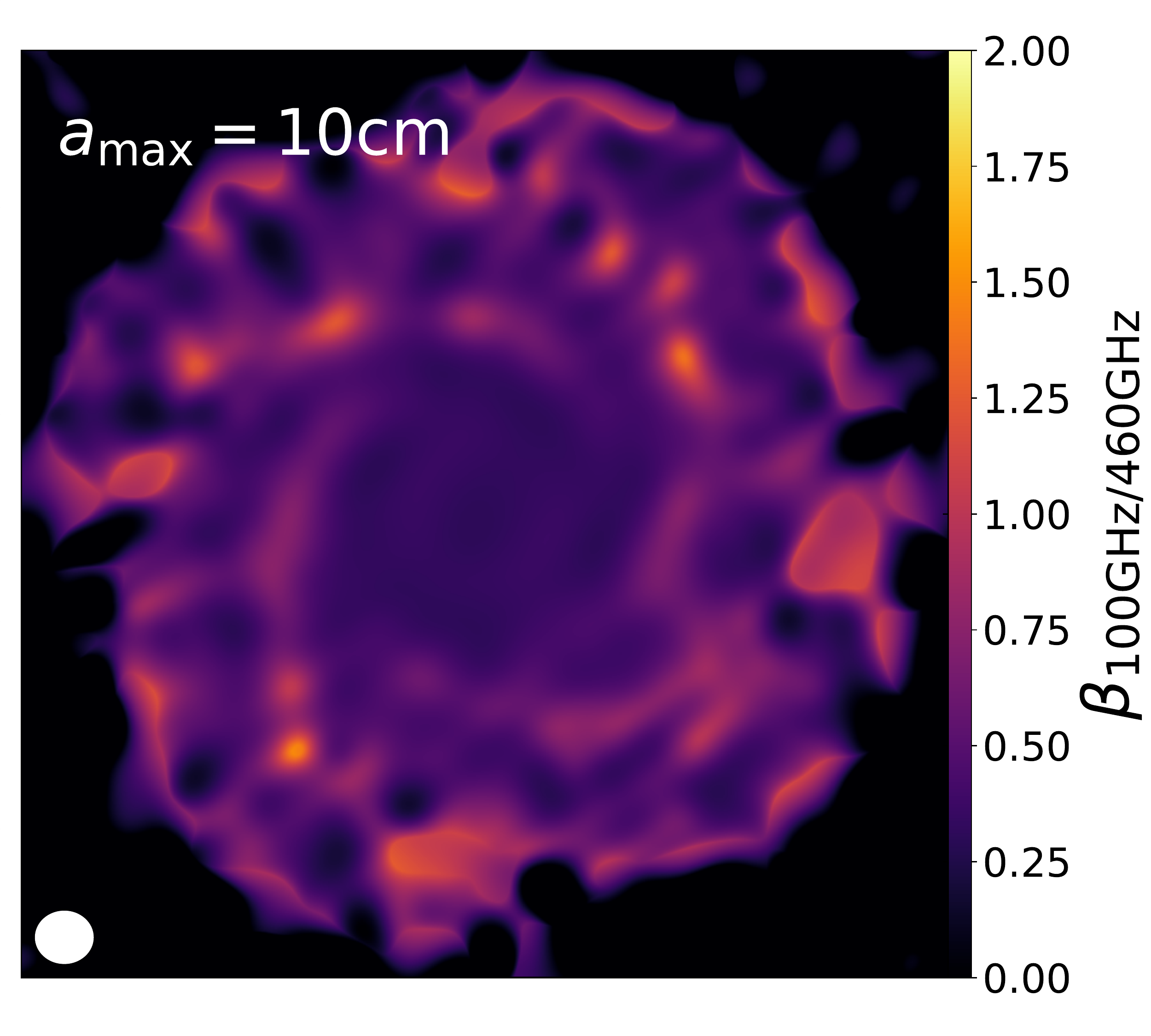}
    \caption{\label{fig:beta10cm} Pixelwise opacity spectral index, $\beta$, derived from synthetic observations of discs with $\dot{M}=5\times10^{-7}$\,M$_{\odot}$yr$^{-1}$ and grain size distributions $n(a)\propto a^{-3.5}$ where $a_{\rm max}=1$\,mm (Left) and $a_{\rm max}=10$\,cm (Right).}
\end{figure*}

Multi-wavelength observations of discs allow us to probe grain growth and dust trapping through calculation of the disc opacity spectral index, $\beta$ \citep{dipierro15}. In the Rayleigh-Jeans limit of an optically thin disc the dust opacity at sub-mm wavelengths will approximately scale as $\kappa \propto \nu^\beta$, where for interstellar dust grains $\beta_{\rm ism}\approx1.7$. Observations of discs show $\beta_{\rm disc} < \beta_{\rm ism}$ \citep[e.g.][]{testietal03,riccietal10} which can be naturally accounted for by the presence of larger grains in the disc and therefore grain growth \citep{draine06}. In Figure \ref{fig:beta10cm} we calculate the $\beta$-parameter from our synthetic ALMA observations, considering fluxes $\nu_1=460$\,GHz and $\nu_2=100$\,GHz, and discs with $\dot{M}=5\times10^{-7}$\,M$_{\odot}$yr$^{-1}$ and $a_{\rm max}=1$\,mm and $10$\,cm. The pixelwise $\beta$ can be calculated as,
\begin{equation}\label{eq:beta}
    \beta = \frac{{\rm ln}F_1 - {\rm ln}F_2}{{\rm ln}\nu_1 - {\rm ln}\nu_2} - 2,
\end{equation}
where $F_1$ and $F_2$ are the pixelwise fluxes at frequencies $\nu_1$ and $\nu_2$ respectively. Spiral regions display the lowest $\beta$ values due to dust trapping of larger grains, whilst depletion of these same grains in inter-spiral regions produces comparatively higher $\beta$ values. Inner disc regions are optically thick and consequently also display low $\beta$ values. We calculate mean $\beta$-values 1.197 and 0.525 for $a_{\rm max}=1$\,mm and $a_{\rm max}=10$\,cm respectively, where the higher $\beta$ value is consequence of less grain growth in the $a_{\rm max}=1$\,mm disc. Note that both of these discs display $\beta_{\rm disc} < \beta_{\rm ism}$.

Through calculation of the $\beta-$parameter in our disc model we therefore demonstrate how it is possible to retrieve information about the underlying grain distribution in discs, and how our model may be used to probe grain properties in discs which have been observed at multiple wavelengths.

\subsection{Observing self-gravitating discs in Taurus}

We now wish to make observational predictions of self-gravitating discs, considering those at a distance $d\sim140$\,pc comparable to the Taurus star-forming region. We setup a suite of discs as described in Section \ref{sec:discparams} and refer the reader to the unsharp masked disc images presented in Appendix \ref{appendix:gallery} for this discussion.

\begin{figure*}
    \centering
    \includegraphics[height=53mm]{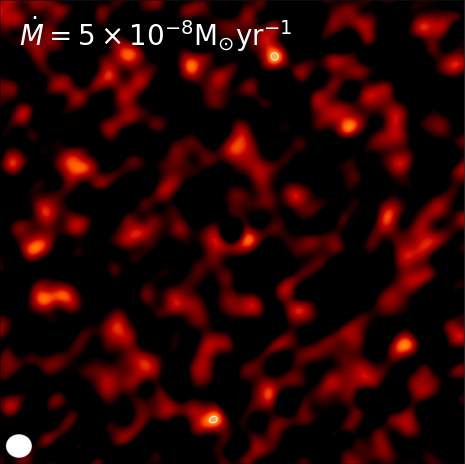}
    \includegraphics[height=53mm]{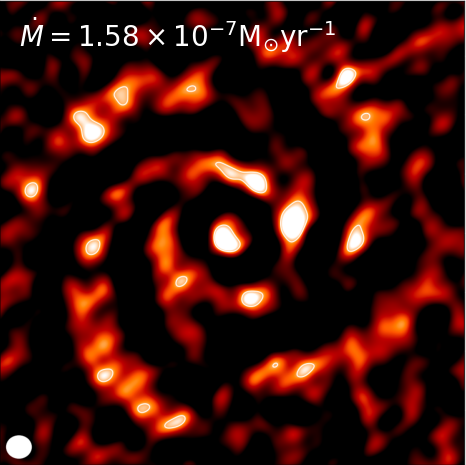}
    \includegraphics[height=53mm]{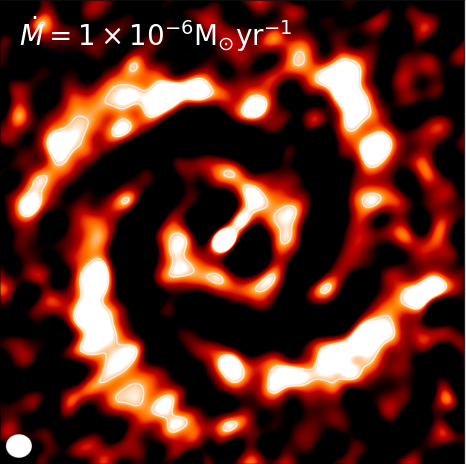}
    \includegraphics[height=53mm]{cbar_115GHz.png}
    \caption{\label{fig:discsvs.mdot} Unsharp masked disc images observed at $f_{\rm obs}=115$\,GHz ($\lambda=2.6$\,mm) in \textsc{casa}. Each disc model has maximum grain size $a_{\rm max}=1$\,mm, $R_{\rm out}=100$\,AU and $\dot{M}=5\times10^{-8}$\,M$_{\odot}$yr$^{-1}$ (Left), $\dot{M}=1.58\times10^{-7}$\,M$_{\odot}$yr$^{-1}$ (Middle), $\dot{M}=1\times10^{-6}$\,M$_{\odot}$yr$^{-1}$ (Right). Observation exposure time, array configuration and PWV level used for these observations are laid out in Table \ref{tab:casa}.}
\end{figure*}

Spiral amplitude in our models increases as $\delta\Sigma/\Sigma \propto \alpha^{-1/2}$ (equation \ref{eq:dsigma/sigma}), hence is an increasing function of accretion rate (see equation \ref{eq:mdot}). This is illustrated in Figure \ref{fig:discsvs.mdot} for discs with $a_{\rm max}=1$\,mm observed at $f_{\rm obs}=115$\,GHz. Low $\dot{M}$ discs generally exhibit no observable substructure for any grain size distribution, whilst the most massive discs tend to be capable of generating detectable spirals at all frequencies considered here. This does however depend on how much grain growth has occurred, as we require that the dust mass budget is dominated by millimetre/centimetre grains ($a_{\rm max}=\rm mm-cm$ sizes) if we are to resolve any spirals. 

\begin{figure*}
    \centering
    \includegraphics[height=40mm]{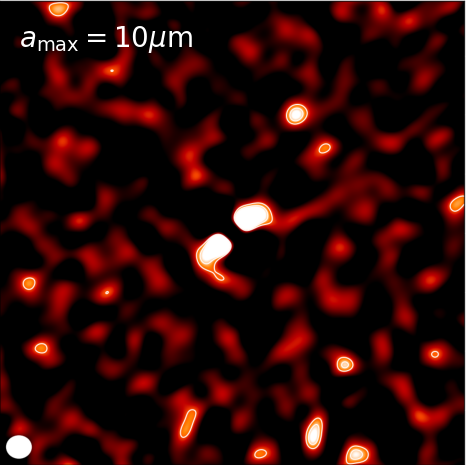}
    \includegraphics[height=40mm]{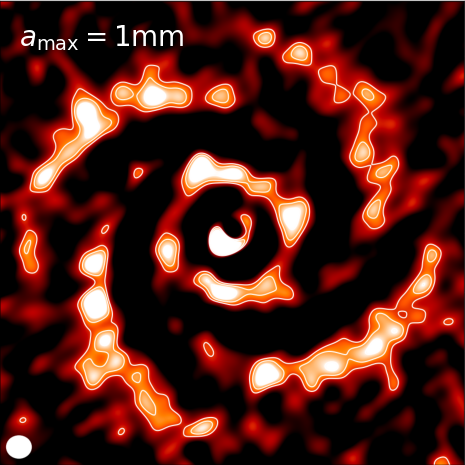}
    \includegraphics[height=40mm]{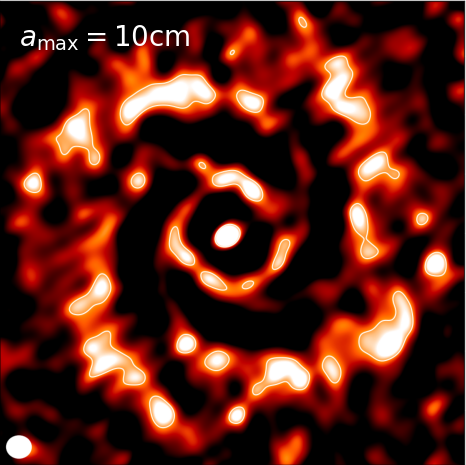}
    \includegraphics[height=40mm]{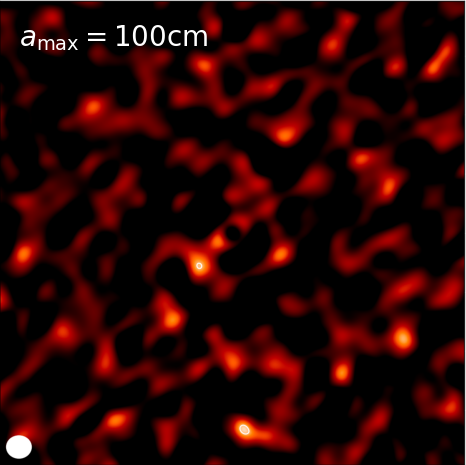}
    \includegraphics[height=40mm]{cbar_115GHz.png}
    \caption{\label{fig:discvs.amax} Unsharp masked disc images observed at $f_{\rm obs}=115$\,GHz ($\lambda=2.6$\,mm) in \textsc{casa}. Each disc has $\dot{M}=5\times10^{-7}$\,M$_{\odot}$yr$^{-1}$, $R_{\rm out}=100$\,AU and we vary $a_{\rm max}$ in the grain size distributions as $10$\,$\mu$m (Left), $1$\,mm (Left middle), $10$\,cm (Right middle) and $100$\,cm (Right). Observation exposure time, array configuration and PWV level used for these observations are laid out in Table \ref{tab:casa}.}
\end{figure*}

Dust emissivity peaks for $\lambda \approx 2\pi a$ \citep{armitage09}, therefore emission from millimetre grains will peak at $\approx$\,millimetre wavelengths. The corresponding wavelengths to the observing frequencies considered here are 2.6mm, 1.3mm and 0.4mm for frequencies of 115GHz, 230GHz and 690GHz respectively. When the dust mass budget is dominated by micron grains or metre-sized objects (i.e. $a_{\rm max}=10$\,${\rm \mu m}$ or $a_{\rm max}=100$\,cm) disc substructure becomes invisible at the ALMA bands considered here as the arm-interarm contrast is low. We illustrate this in Figure \ref{fig:discvs.amax} which shows how emission from spiral regions varies with grain size distribution in discs with $\dot{M}=5\times10^{-7}$\,M$_{\odot}$yr$^{-1}$ observed at $f_{\rm obs}=115$\,GHz. Substructure only becomes recognisable in discs with unfavourable grain size distributions when we observe at shorter wavelengths ($f_{\rm obs}=690$\,GHz, $\lambda=0.4$\,mm), but only in the most highly accreting cases.

Without including dust trapping in their model, \cite{halletal16} previously found a narrow region of parameter space within which self-gravitating discs would display spirals observable with ALMA. They predicted a $100$\,AU disc must be accreting in the range $1\times10^{-7}$\,M$_{\odot}$yr$^{-1} \lesssim \dot{M} \lesssim 1\times10^{-6}$\,M$_{\odot}$yr$^{-1}$, where the maximum accretion rate here is set by the limit at which discs become susceptible to fragmentation. We suggest that in fact spiral emission may be distinct for lower accretion rates than previously predicted, if sufficient grain growth has occurred. The discs in Figures \ref{fig:230gallery} and \ref{fig:690gallery} observed at $230$\,GHz and $690$\,GHz respectively continue to display detectable spiral structure down to the lowest $\dot{M}$ considered here, as long as the dust mass budget is dominated by millimetre/centimetre grains. Note however that we are observing these discs face-on and therefore in favourable conditions for resolving spiral features. Inclining and rotating these discs may well obscure them. However, we would still expect to be able to detect spirals to lower $\dot{M}$ than previously suggested.

It is also intriguing that we calculate the fragmentation threshold to fall almost exactly coincident with the ideal $a_{\rm max}$ values for detecting spirals (see Figures \ref{fig:afrag10} and \ref{fig:afrag30}). We should therefore not be surprised if we find that in fact the grain size distributions of self-gravitating discs fall within this ideal region of parameter space.

\section{Analysing discs from the DSHARP sample}\label{sec:dsharp}

\begin{figure*}
    \centering
    \includegraphics[width=.3\textwidth]{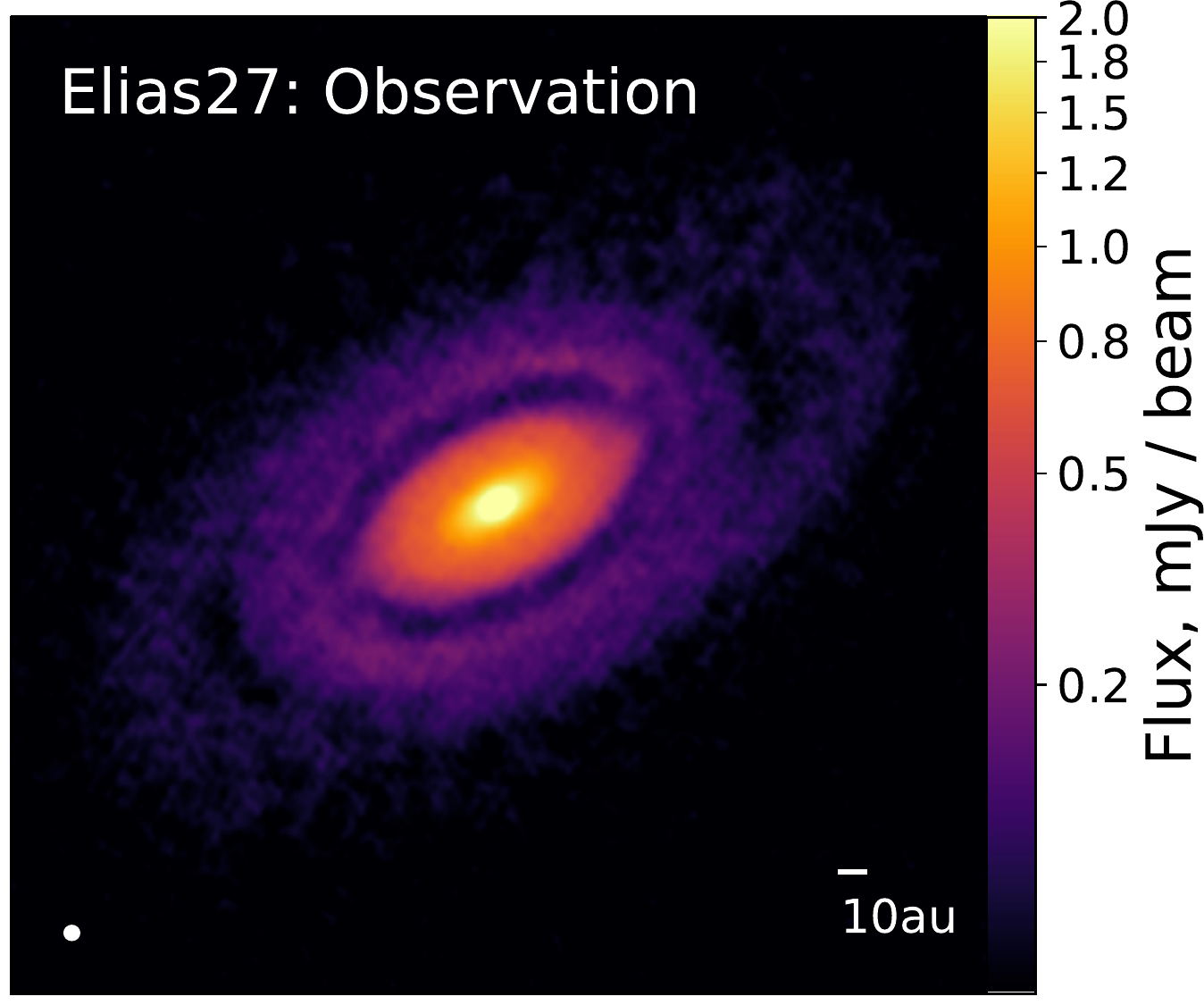}
    \includegraphics[width=.3\textwidth]{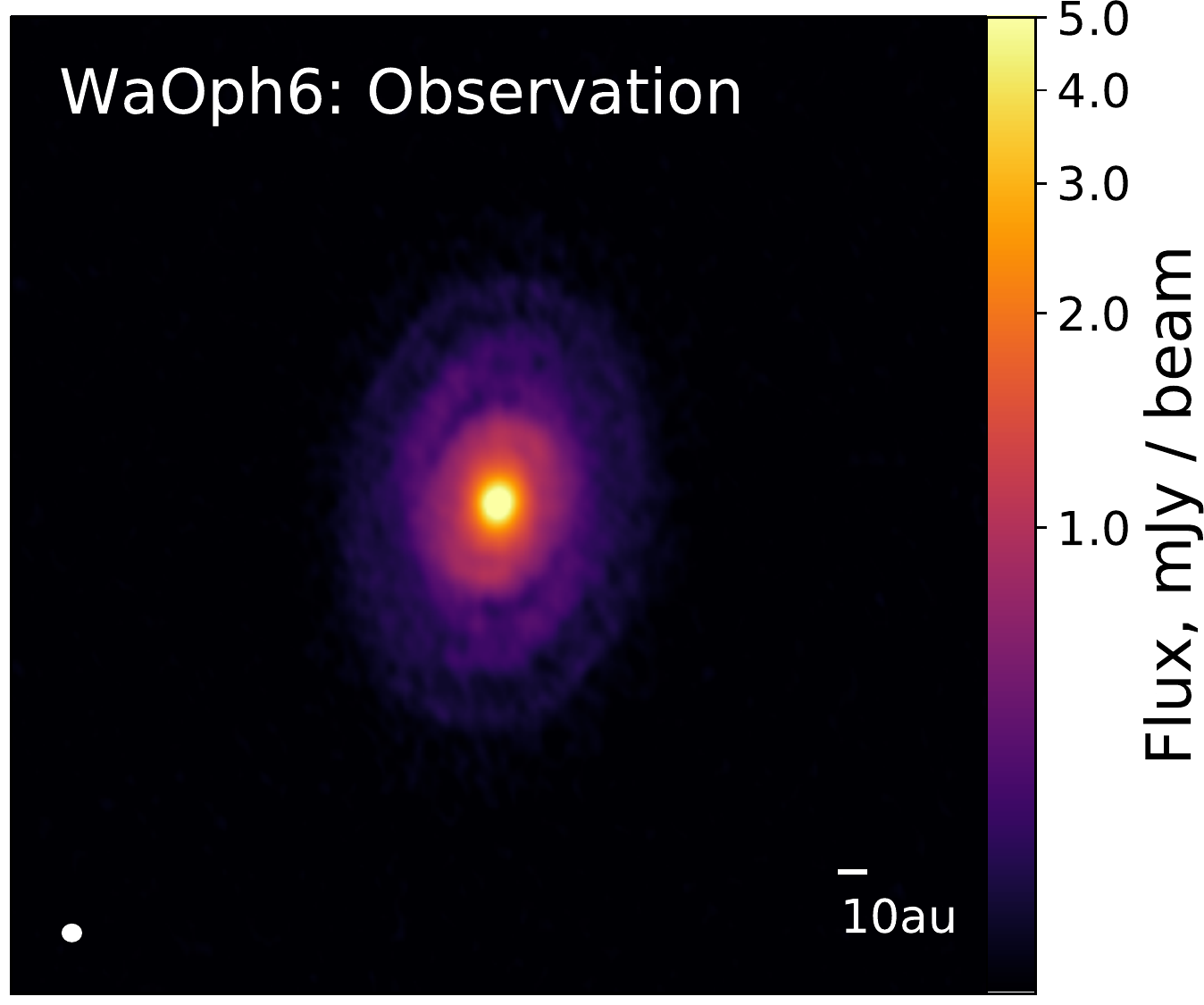}
    \includegraphics[width=.3\textwidth]{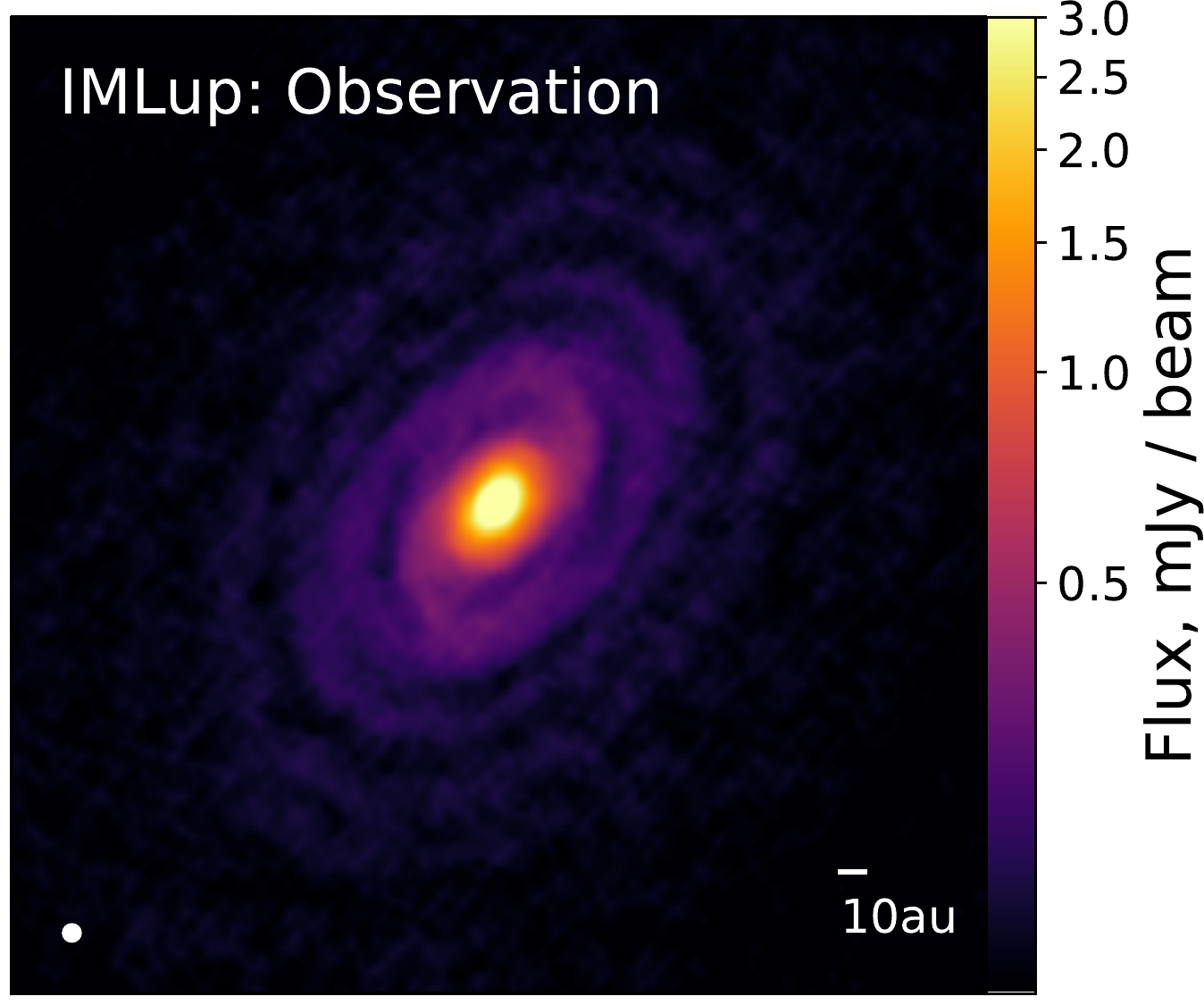}
    \caption{\label{fig:dsharp} ALMA 240GHz (1.3mm) continuum images of Elias 27 (Left), WaOph 6 (Middle) and IM Lup (Right) \citep{dsharp1}. ALMA antenna configurations, observing times and PWV levels for these observations are laid out in Table \ref{tab:dsharpcasa}.}
\end{figure*}

We now turn our model to analysing real observational data of potential self-gravitating discs. The recent DSHARP survey studied 20 nearby protoplanetary discs using ALMA, with 3 of these discs exhibiting spiral substructure \citep{dsharp1,dsharp3}. The ALMA continuum images from this survey of the Elias 27, WaOph 6 and IM Lup discs are shown in Figure \ref{fig:dsharp}.

We use our models to investigate if the observed substructure in these 3 systems can be explained through the gravitational instability, or if instead they require an alternative explanation.

Although well within the capability of our models, a complete examination of the potential parameter space of these discs is beyond the scope of the work presented here. Instead, we simply model these 3 systems using the disc parameters derived in \cite{dsharp1} and \cite{dsharp3}, and make predictions as to whether we should expect these systems to produce self-gravitating spiral substructure observable with ALMA. The disc parameters used are laid out in Table \ref{tab:dsharp}. We setup these discs with dust size distribution $
n(a) \propto a^{-3.5}$, with $a_{\rm min}=0.1\mu$m and set $a_{\rm max}$ as the fragmentation threshold where $v_{\rm frag}=10$\,ms$^{-1}$ (equation \ref{eq:afrag}), and use the canonical dust-to-gas ratio of 0.01.

Residual images in \cite{dsharp3} are produced by deprojecting the discs and subtracting their median axisymmetric radial profiles. We do the same here by binning each disc into $1$\,AU-wide radial bins and subtracting the median azimuthal fluxes. We re-derive the residual images for each of the original DSHARP observations in this way, as well as for our disc models. For each disc observation and model, we show deprojected continuum and residual images (with ${\rm PA}=0^{\circ}$ and  $i=0^{\circ}$), presenting our results in Figures \ref{fig:elias}, \ref{fig:waoph} and \ref{fig:imlup}. In each case we provide reference colorbars for direct comparison between the fluxes of the disc models and observations, and each disc model and counterpart observation is plotted between the same flux range for ease of comparison.

Logarithmic spiral structure is imposed in each disc model using values of $a$ and $b$ (equation \ref{eq:spiraltheta}) derived by \cite{dsharp3}. Best-fit values of $a$ and $b$ that we find from those quoted in \cite{dsharp3} are laid out in Table \ref{tab:dsharp}.

We produce synthetic observations of each disc using \textsc{casa} with observing setups consistent with those outlined in \cite{dsharp1}. We observe each disc for $t_{\rm obs}=3600$\,s using array configuration C40-8. For each observation we use PWV values at the upper bound of the quoted range in \cite{dsharp1}, setting values 1.35mm, 1.30mm and 1.05mm for Elias 27, IM Lup and WaOph 6 respectively. Input parameters for \textsc{casa} used for each disc are laid out in Table \ref{tab:dsharpcasa}.

\begin{table*}
\centering
\begin{tabular}{|| c || c c c c c c c c c c ||}
\hline
Disc & log$M_*$ $^{[1]}$ & $R_*$ & log$T_{\rm eff}$ $^{[1]}$ & $R_{\rm spirals}$ $^{[2]}$ & log$\dot{M}$ $^{[1]}$ & d $^{[2]}$ & $i$ $^{[2]}$ & PA $^{[2]}$ & a $^{[2]}$ & b $^{[2]}$ \\
 & (M$_{\odot}$) & (AU) & (K) & (AU) & (M$_{\odot}$\,yr$^{-1}$) & (pc) & ($^{\circ}$) & ($^{\circ}$) & (AU) & \\
 \\
 (1) & (2) & (3) & (4) & (5) & (6) & (7) & (8) & (9) & (10) & (11) \\
\hline\hline
\\
 Elias 27 & $-0.31^{+0.15}_{-0.11}$ & 2.3 & $3.59\pm0.03$ & 50-230 & $-7.2\pm0.5$ & 116$^{+19}_{-10}$ & 56.2 & 118.8 & 110.9 & -0.282 \\
 \\
 WaOph 6 & $-0.17^{+0.17}_{-0.09}$ & 3.2 & $3.62\pm0.03$ & 25-75 & $-6.6\pm0.5$ & 123$\pm$2 & 47.3 & 174.2 & 45.9 & 0.238 \\  
 \\
 IM Lup & $-0.05^{+0.09}_{-0.13}$ & 2.5 & $3.63\pm0.03$ & 25-110 & $-7.9\pm0.4$ & 158$\pm$3 & 47.5  & 144.5 & 43 & -0.181 \\ 
 \\
 \hline
\end{tabular}
\caption{\label{tab:dsharp} Disc model parameters used in our modelling of the DSHARP discs in Section \ref{sec:dsharp}.  Columns are as follows. (1) Disc being modelled. (2) Log stellar mass. (3) Stellar radius. (4) Log effective temperature of the star. (5) Spiral inner and outer radii considered here. (6) Log mass accretion rate. (7) Distance to the system. (8) Disc inclination. (9) Disc position angle. (10) Best-fit logarithmic spiral a (Equation \ref{eq:spiraltheta}). (11) Best-fit logarithmic spiral  b (Equation \ref{eq:spiraltheta}).}
\end{table*}

\begin{table}
\centering
\begin{tabular}{|| c c c c c ||}
\hline
Disc & $f_{\rm obs}$ & $t_{\rm obs}$ & \parbox{1cm}{\centering Antenna \\ Config} & PWV Level \\
& (1) & (2) & (3) & (4) \\
\hline\hline
\\
 Elias 27 & 240\,GHz & 3600\,s & C40-8 & 1.35\,mm \\
 \\
 WaOph 6 & 240\,GHz & 3600\,s & C40-8 & 1.30\,mm \\  
 \\
 IM Lup & 240\,GHz & 3600\,s & C40-8 & 1.05\,mm \\ 
 \\
 \hline
\end{tabular}
\caption{\label{tab:dsharpcasa} Input parameters used here for generating synthetic images with \textsc{casa} for the modelled DSHARP discs. (1) ALMA observing frequency. (2) Simulated observing time. (3) ALMA antenna configuration used. (4) Precipitable Water Vapour (PWV) level.}
\end{table}

\subsection{Elias 27}

\begin{figure}
    \centering
    \includegraphics[width=.49\linewidth]{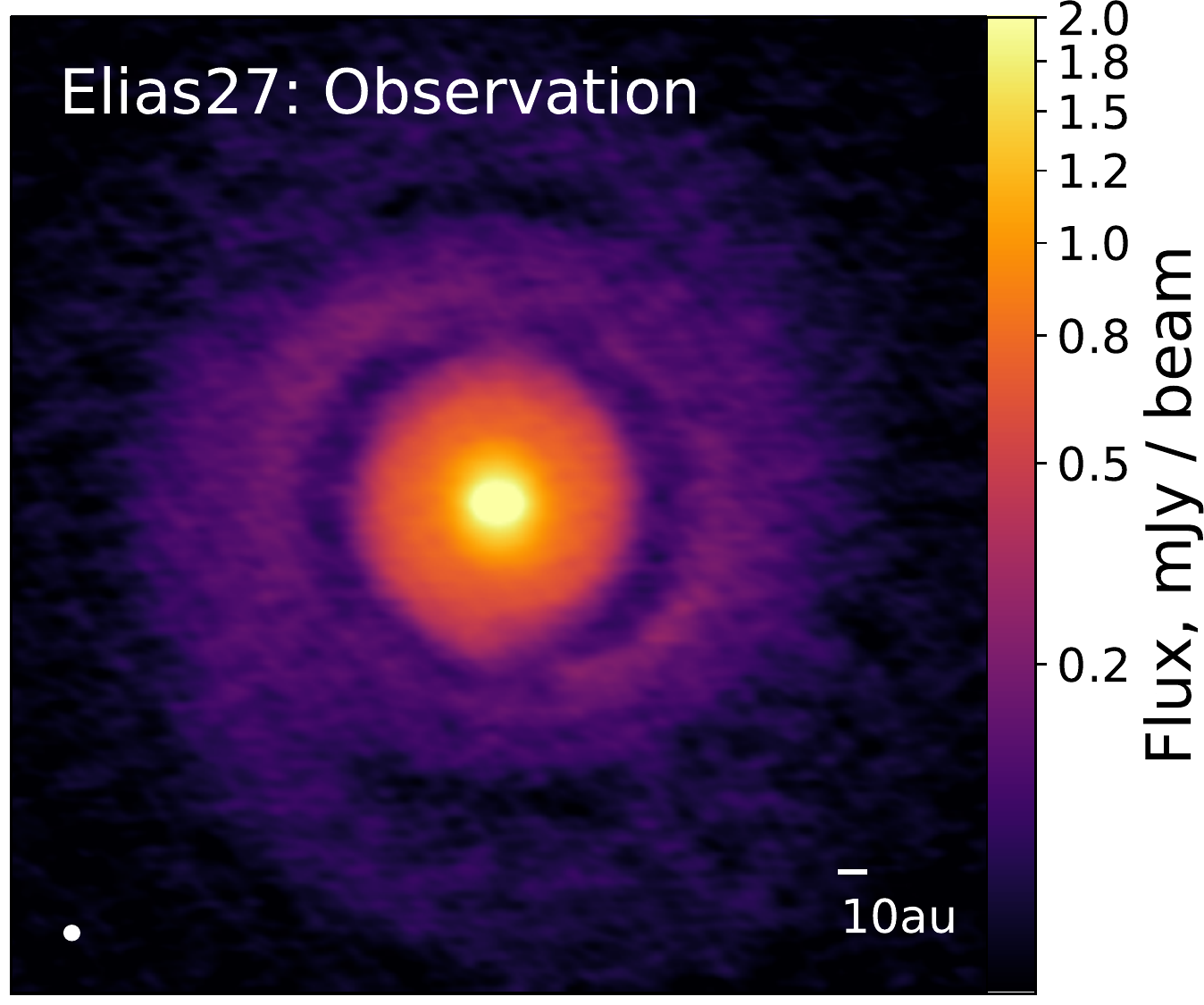}
    \includegraphics[width=.49\linewidth]{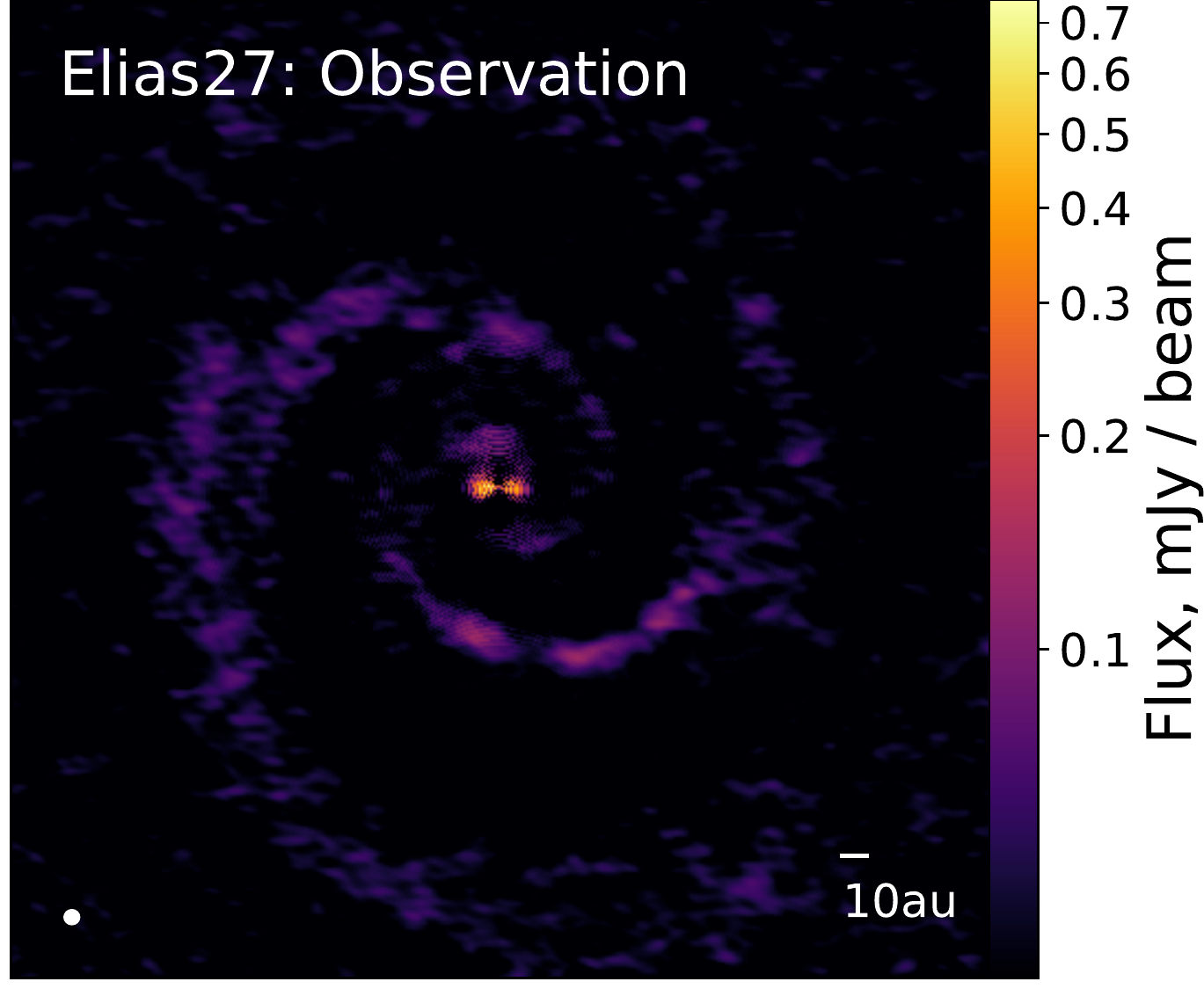}
    \includegraphics[width=.49\linewidth]{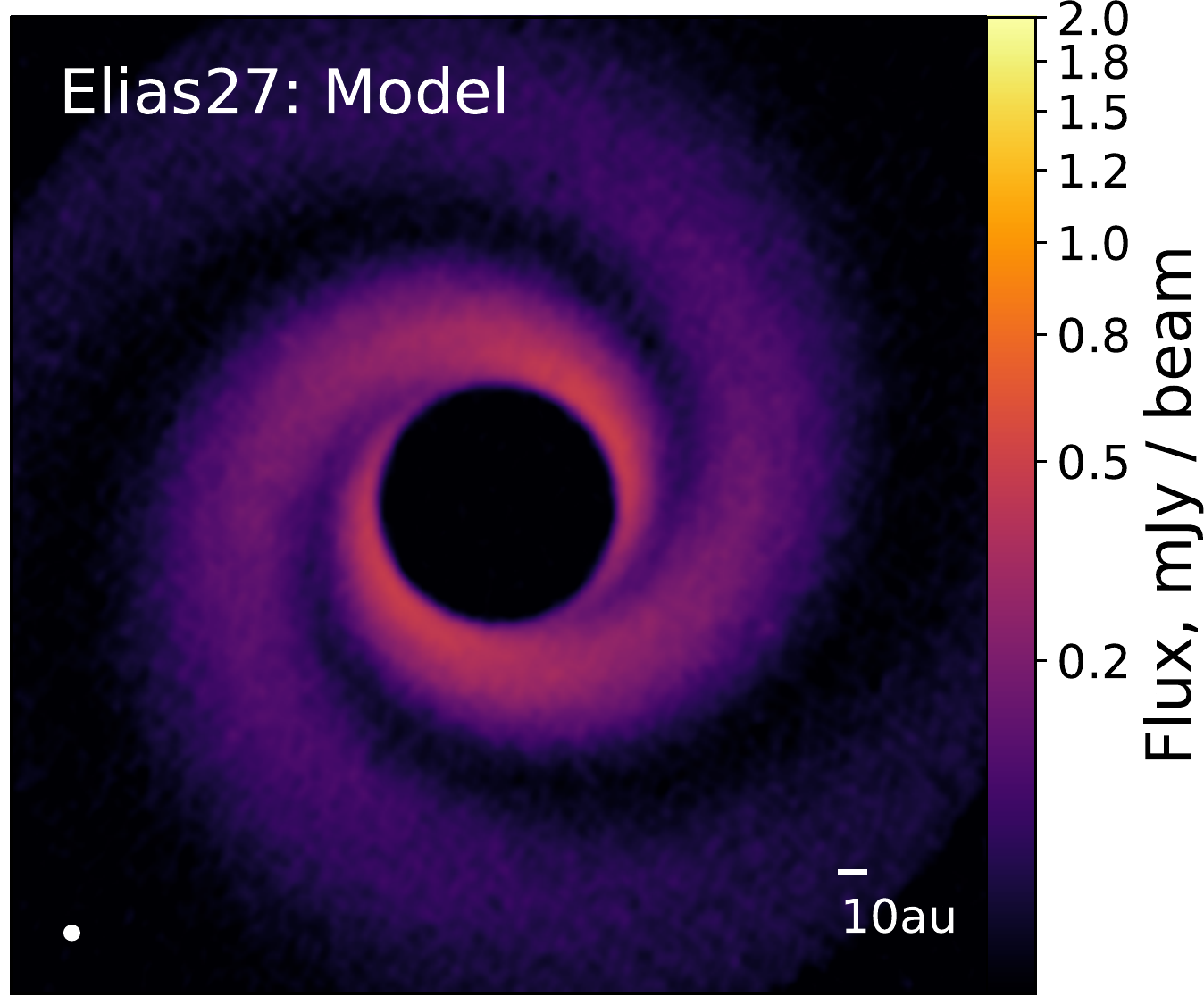}
    \includegraphics[width=.49\linewidth]{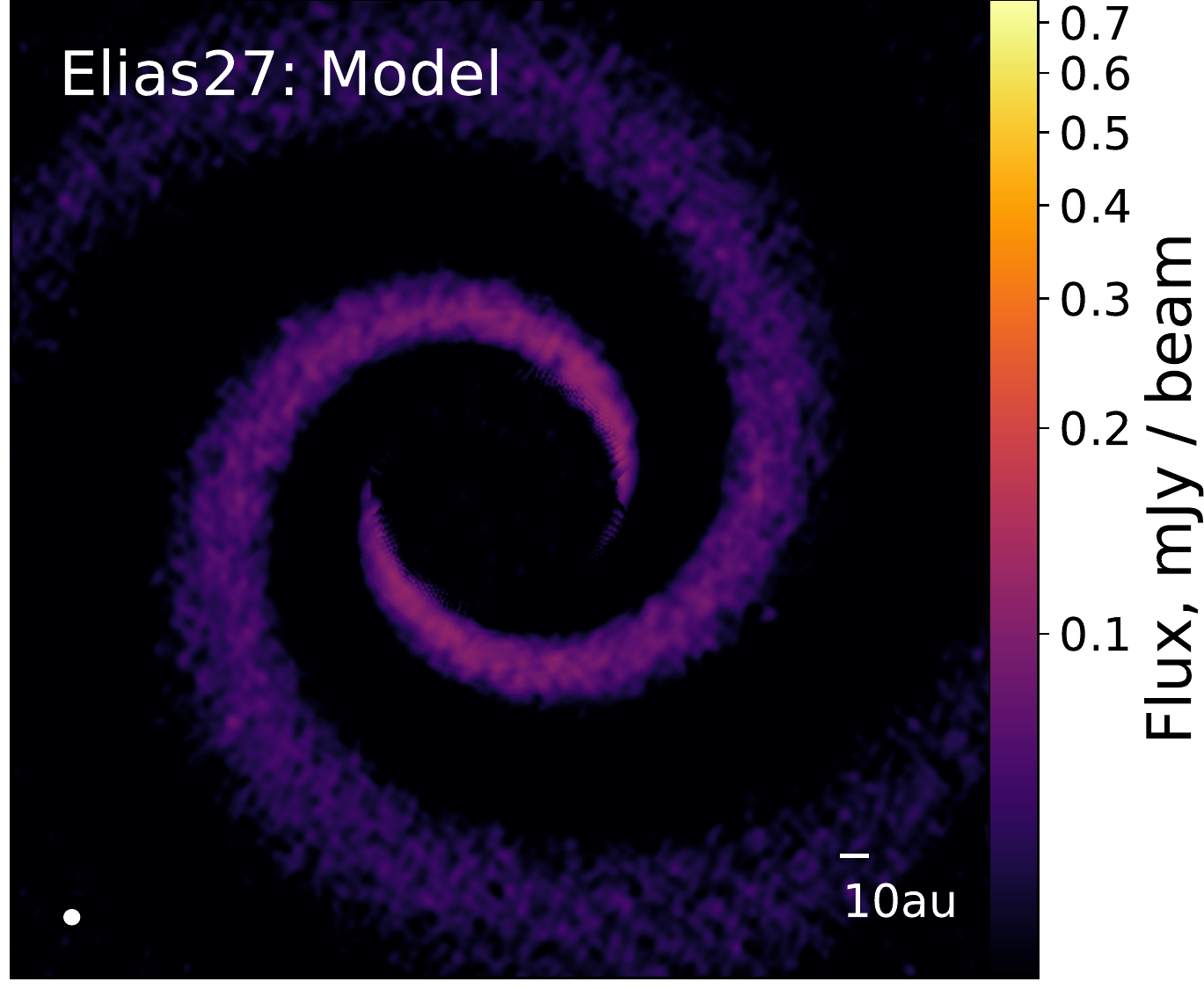}
    \caption{\label{fig:elias} Elias 27 discs images. Top: Deprojected ALMA continuum image (left) and residual profile (right). Bottom: Deprojected disc model continuum image (left) and residual profile (right). Input properties for the disc models and observation parameters are laid out in Tables \ref{tab:dsharp} and \ref{tab:dsharpcasa}.}
\end{figure}

Elias 27 is a 0.8\,Myr M0 star located in the $\rho$ Oph star forming region at a distance $d=116^{+19}_{-10}$\,pc \citep{gaiaDR2,dsharp1}. The residual profile of the Elias 27 continuum image (Figure \ref{fig:elias}) shows two symmetric spiral arms extending from $R_{\rm in}\sim50$\,AU to $R_{\rm out}\sim230$\,AU, with ${\rm PA}=118.8^{\circ}$ and $i=56.2^{\circ}$ \citep{dsharp3}. 

The spiral structure of Elias 27 is probably the most well-studied of the three discs here. The system has previously been modelled using both both grid-based and SPH simulations, with authors such as \cite{meruetal17}, \cite{tomidaetal17} and \cite{halletal18} all finding GI to be a plausible explanation for the observed morphology. Estimates of the Toomre parameter in the disc however suggest that Elias 27 should be gravitationally stable at all radii \citep{perezetal16}, but this comes with the caveat that estimates of $Q$ are subject to high levels of uncertainty. Further research where the constraints on the disc mass and temperature are improved may lead to different conclusions.

We set up our disc model with ${\rm log}M_* {\rm(M_\odot)}=-0.31$, $R_*=2.3$\,AU, ${\rm log}T_{\rm eff}{\rm(K)}=3.59$ and ${\rm log}\dot{M}{\rm(M_{\odot}yr^{-1})}=-7.2$ \citep{dsharp1, dsharp3}. Logarithmic spiral structure is imposed with $a=76.0$\,AU and $b=-0.29$ extending from $R=50-230$\,AU, where we use a mask to remove the inner $50$\,AU from our observations to avoid the spirals being washed out by the brighter central region.

Our model calculates Elias 27 to have a disc mass $M_{\rm disc}=0.13$\,M$_{\odot}$ inside $R_{\rm out}=230$\,AU, and therefore $q=0.27$. Figure \ref{fig:elias} shows the resultant synthetic observations generated from our models, exhibiting clear self-gravitating spiral structure in both the deprojected continuum and residual images.

\subsection{WaOph 6}

WaOph 6 is a 0.3\,Myr K6 star located in the $\rho$ Oph star forming region at a distance $d=123\pm2$\,pc \citep{gaiaDR2,dsharp1}. After subtracting the axisymmetric radial profile, two compact spiral arms are revealed which extend from $R_{\rm in}\sim25$\,AU to $R_{\rm out}\sim75$\,AU, with ${\rm PA}=174.2^{\circ}$ and $i=47.3^{\circ}$ \citep{dsharp3}. 

In their analysis of the morphology of gravitationally unstable discs, \cite{dongetal15} suggest that for a disc to be gravitationally unstable it must be compact ($R\leq100$\,AU) and highly accreting at a rate $\dot{M}\geq10^{-6}$\,M$_{\odot}$yr$^{-1}$. WaOph 6 has the highest accretion rate and the most compact spiral structure of the 3 discs in question here, both of which are close to matching these suggested criteria.

We setup our disc model with ${\rm log}M_*{\rm(M_\odot)}=-0.17$, $R_*=3.2$\,AU, ${\rm log}T_{\rm eff}{\rm(K)}=3.62$ and ${\rm log}\dot{M}{\rm(M_{\odot}yr^{-1})}=-6.6$ \citep{dsharp1,dsharp3}. Logarithmic spirals are imposed with $a=34.0$\,AU and $b=0.24$ extending from $R=25-75$\,AU, where again we mask the inner $25$\,AU of the disc images.

\begin{figure}
    \centering
    \includegraphics[width=.49\linewidth]{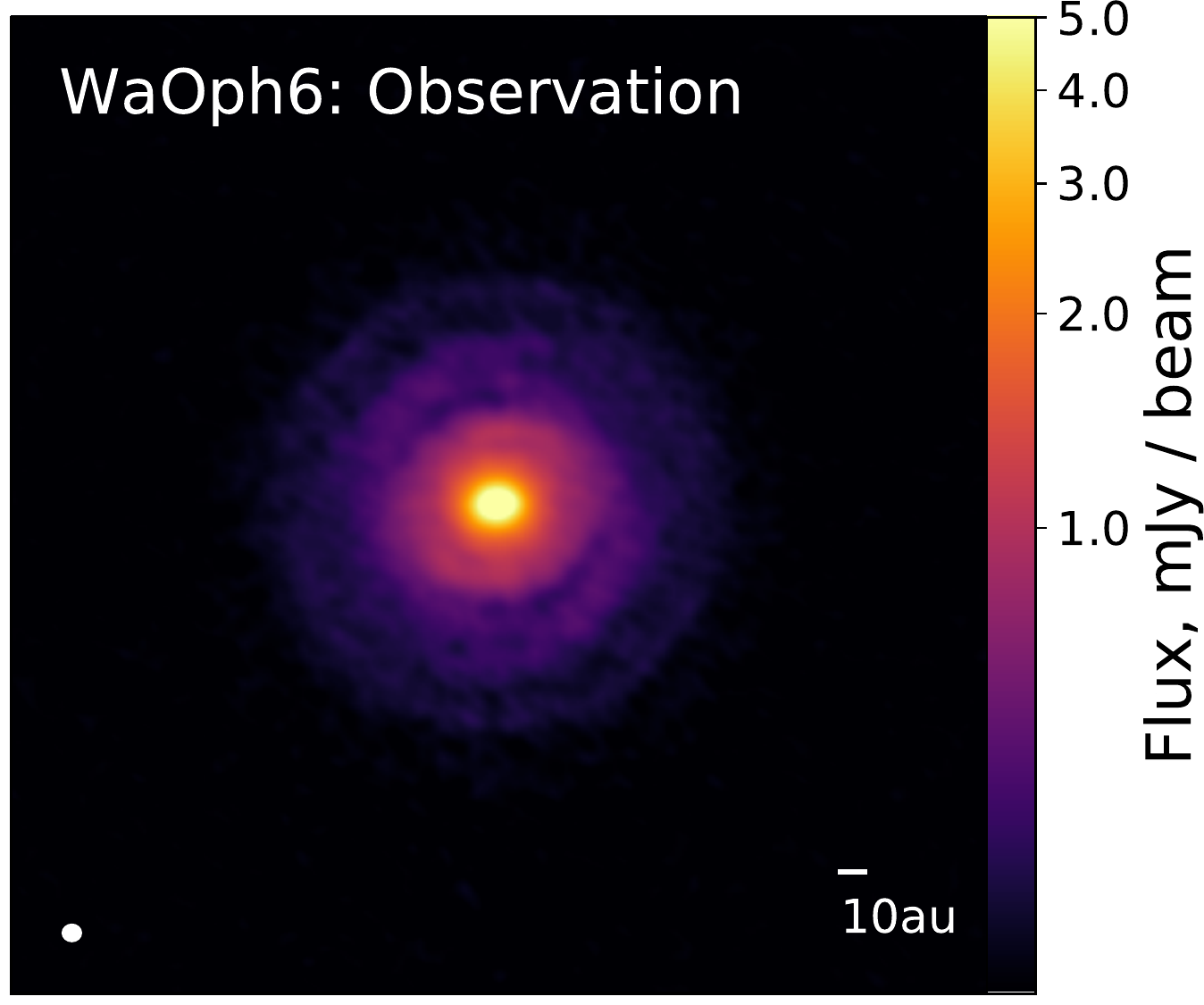}
    \includegraphics[width=.49\linewidth]{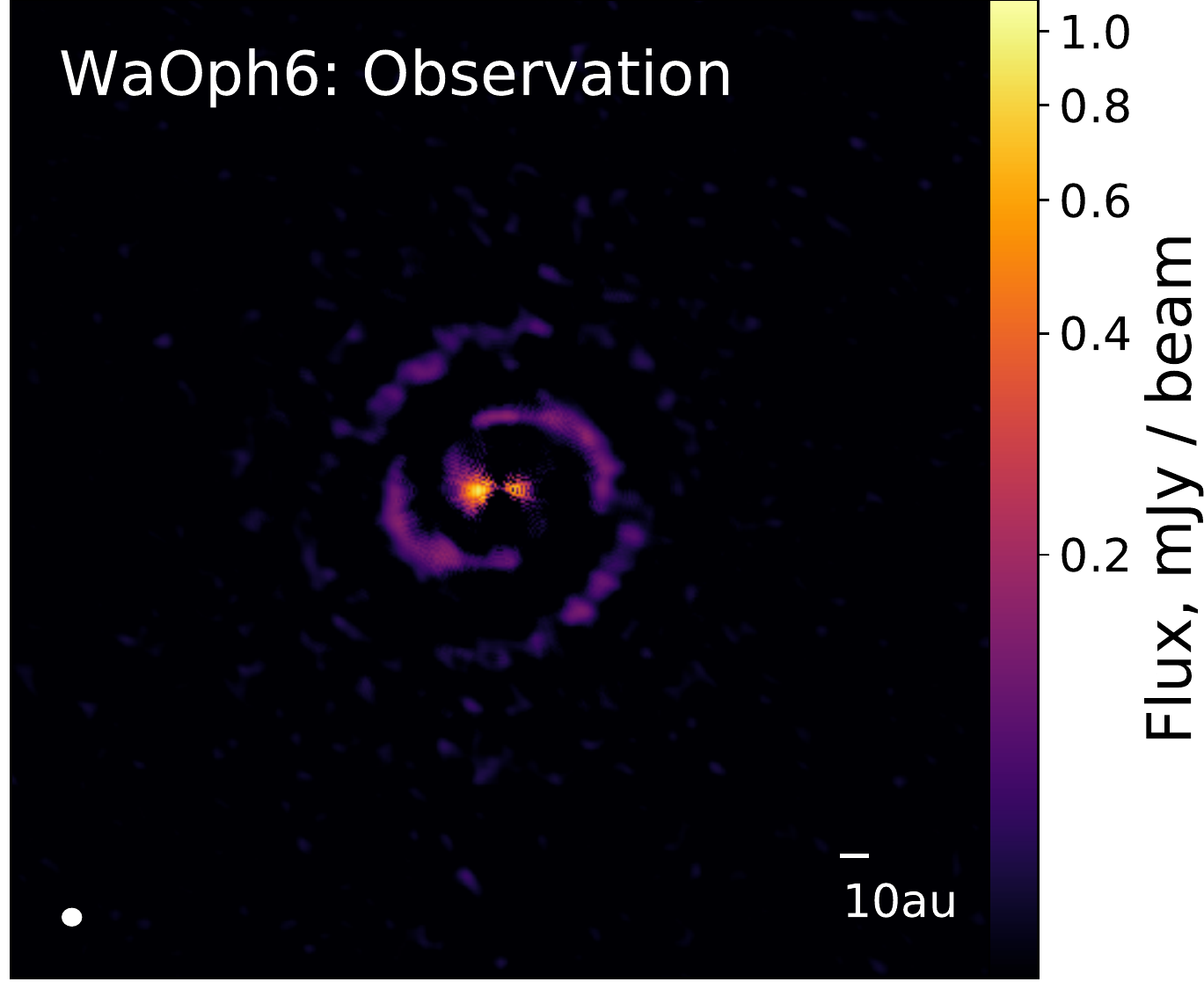}
    \includegraphics[width=.49\linewidth]{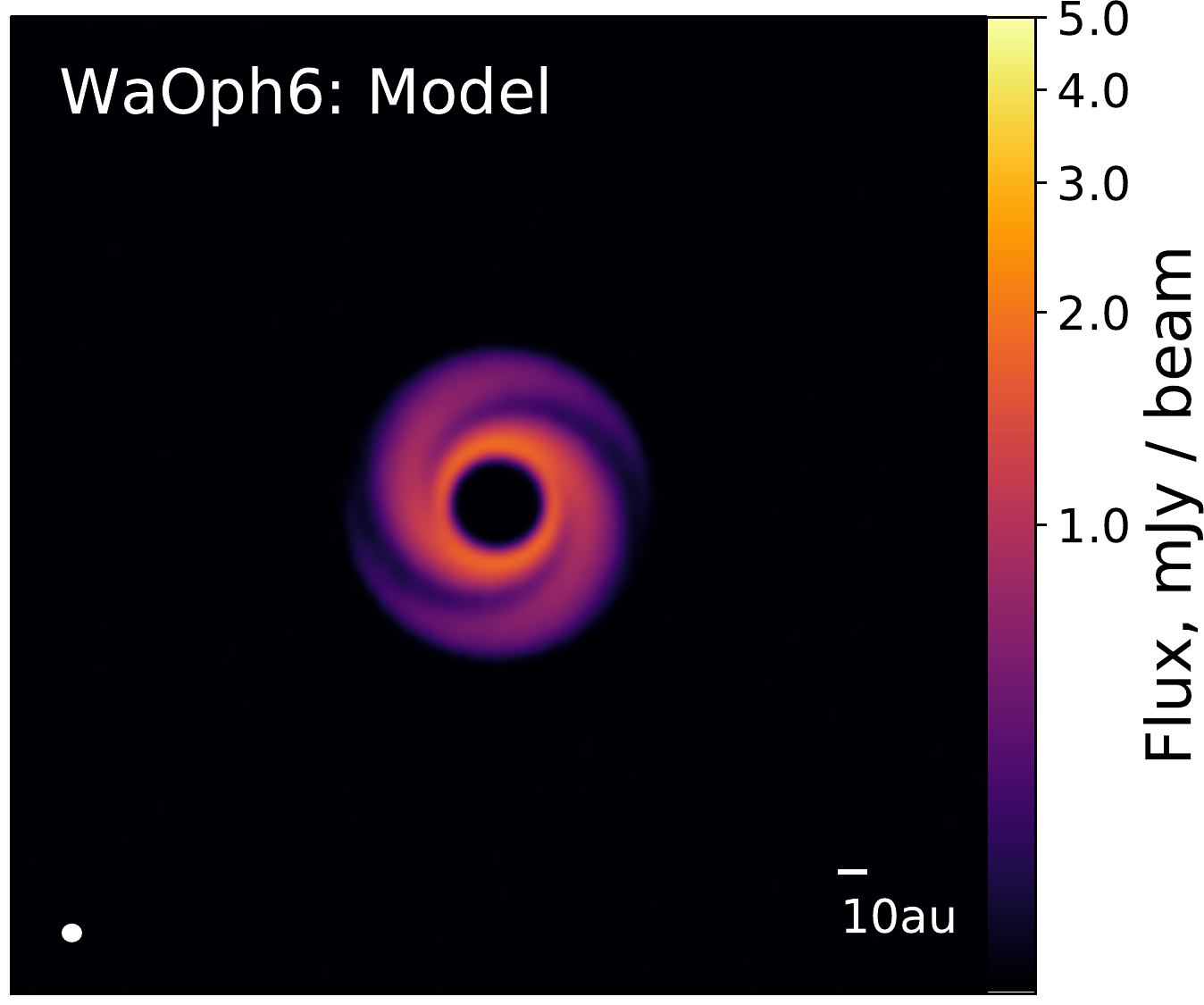}
    \includegraphics[width=.49\linewidth]{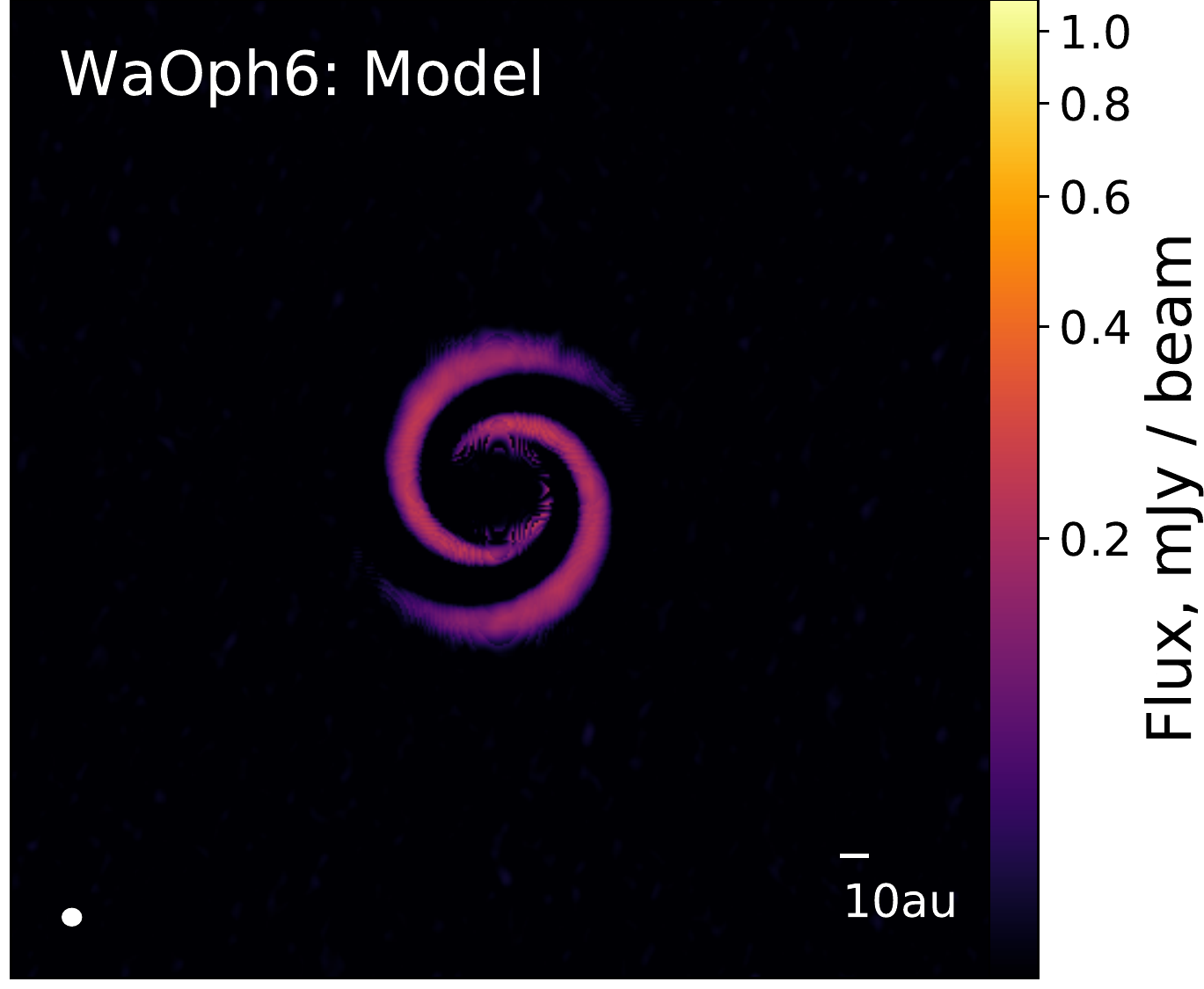}
    \caption{\label{fig:waoph} WaOph 6 discs images. Top: Deprojected ALMA continuum image (left) and residual profile (right). Bottom: Deprojected disc model continuum image (left) and residual profile (right). Input properties for the disc models and observation parameters are laid out in Tables \ref{tab:dsharp} and \ref{tab:dsharpcasa}.}
\end{figure}

We calculate WaOph 6 to have a disc mass $M_{\rm disc}=0.16$\,M$_{\odot}$ and therefore $q=0.24$. Our models reproduce distinct observable, self-gravitating spiral structure in both the deprojected continuum and residual images shown in Figure \ref{fig:waoph}. 

\subsection{IM Lup}

IM Lup is a 0.5Myr K5 star in the Lupus II cloud at a distance $d=158\pm3$\,pc \citep{gaiaDR2,dsharp1}. Residual profiles of the IM Lup continuum images reveal two spirals extending from $R_{\rm in}\sim25$\,AU to $R_{\rm out}=110$\,AU, with PA$=144.5^{\circ}$ and $i=47.5^{\circ}$ \citep{dsharp3}. 

Previous detection of any spiral structure in the IM Lup system has been minimal, with observed substructures being classified as two concentric rings at $R\approx95$\,AU and $R\approx320$\,AU, and only tenuous reports of the possibility of tightly wound spirals \citep{avenhausetal18}. \cite{cleevesetal16} report a massive, gravitationally stable disc with a minimum Toomre parameter $Q_{\rm min}=3.7$ at $R=70$\,AU and an extended CO disc to $R=970$\,AU, making IM Lup one of the largest protoplanetary discs detected to date.

We model the disc here out to $R_{\rm out}=110$\,AU, consistent with the radial extent of the observed spiral structure reported in \cite{dsharp3}. Our disc model is setup with ${\rm log}M_*{\rm(M_{\odot})}=-0.05$, $R_*=2.5$\,AU, ${\rm log}T_{\rm eff}{\rm(K)}$ and ${\rm log}\dot{M}{\rm(M_{\odot}yr^{-1})}=-7.9$ \citep{dsharp1,dsharp3}. We impose logarithmic spiral structure with $a=59.0$\,AU and $b=0.18$ extending from $R=25-110$\,AU.

\begin{figure}
    \centering
    \includegraphics[width=.49\linewidth]{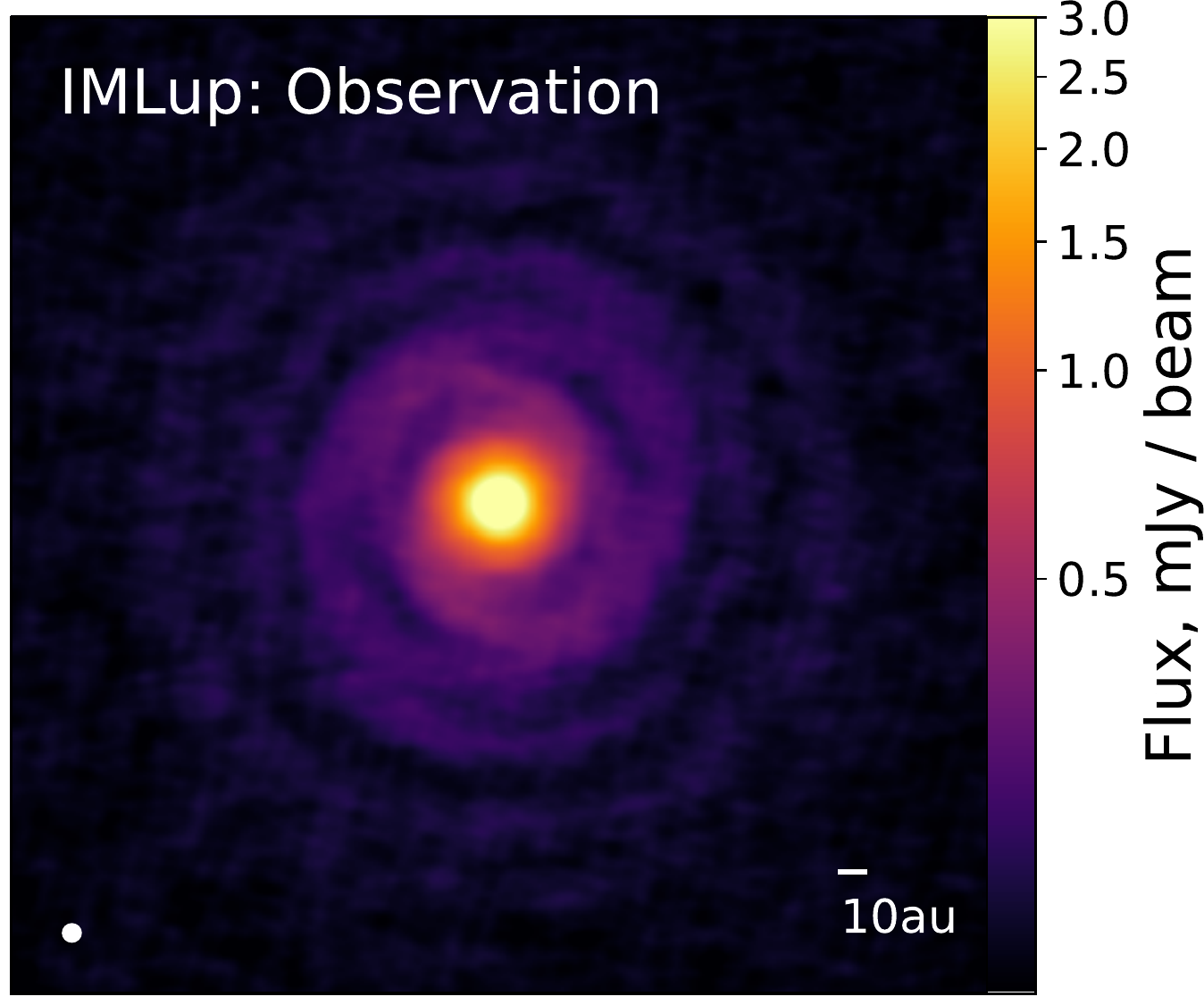}
    \includegraphics[width=.49\linewidth]{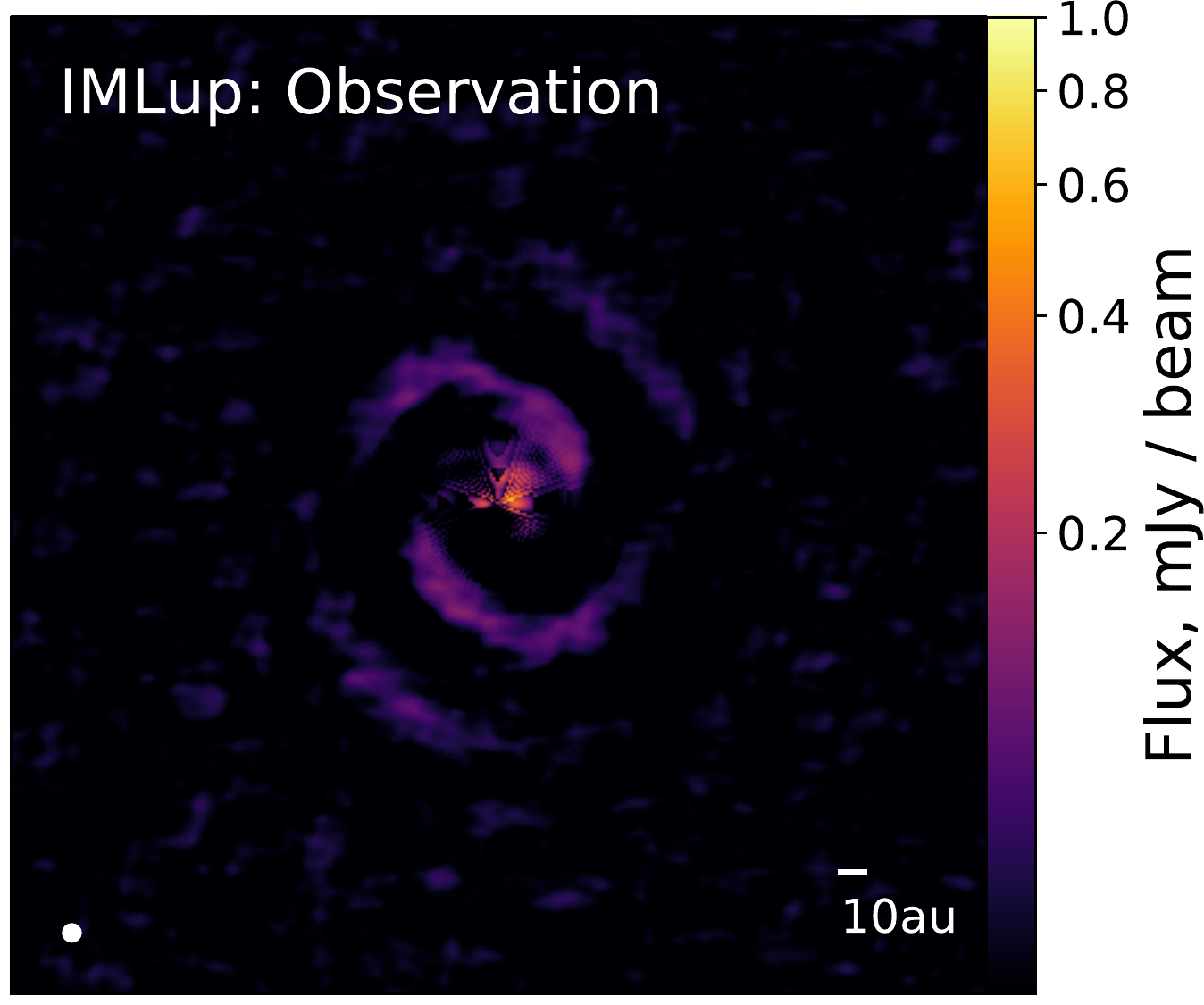}
    \includegraphics[width=.49\linewidth]{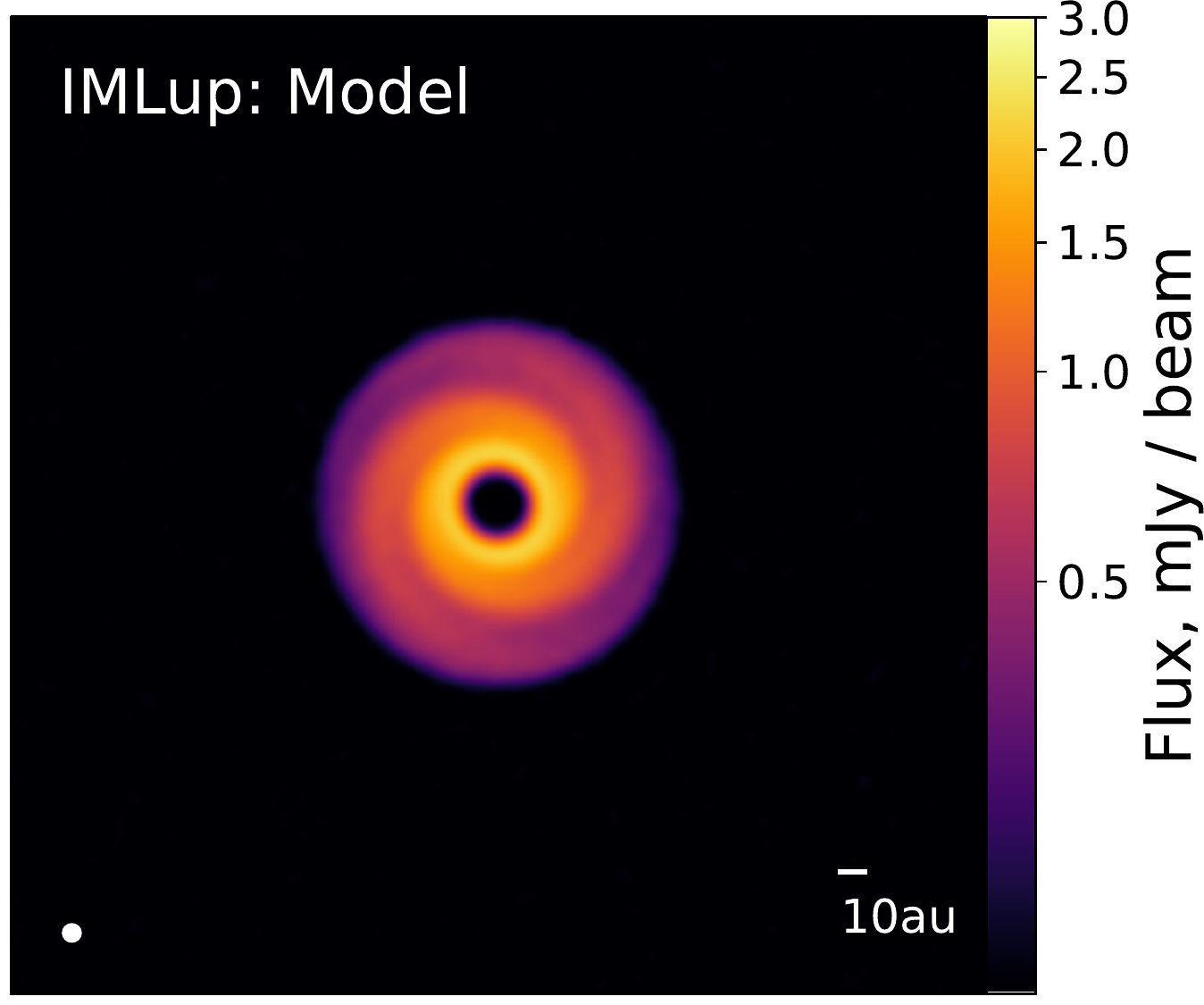}
    \includegraphics[width=.49\linewidth]{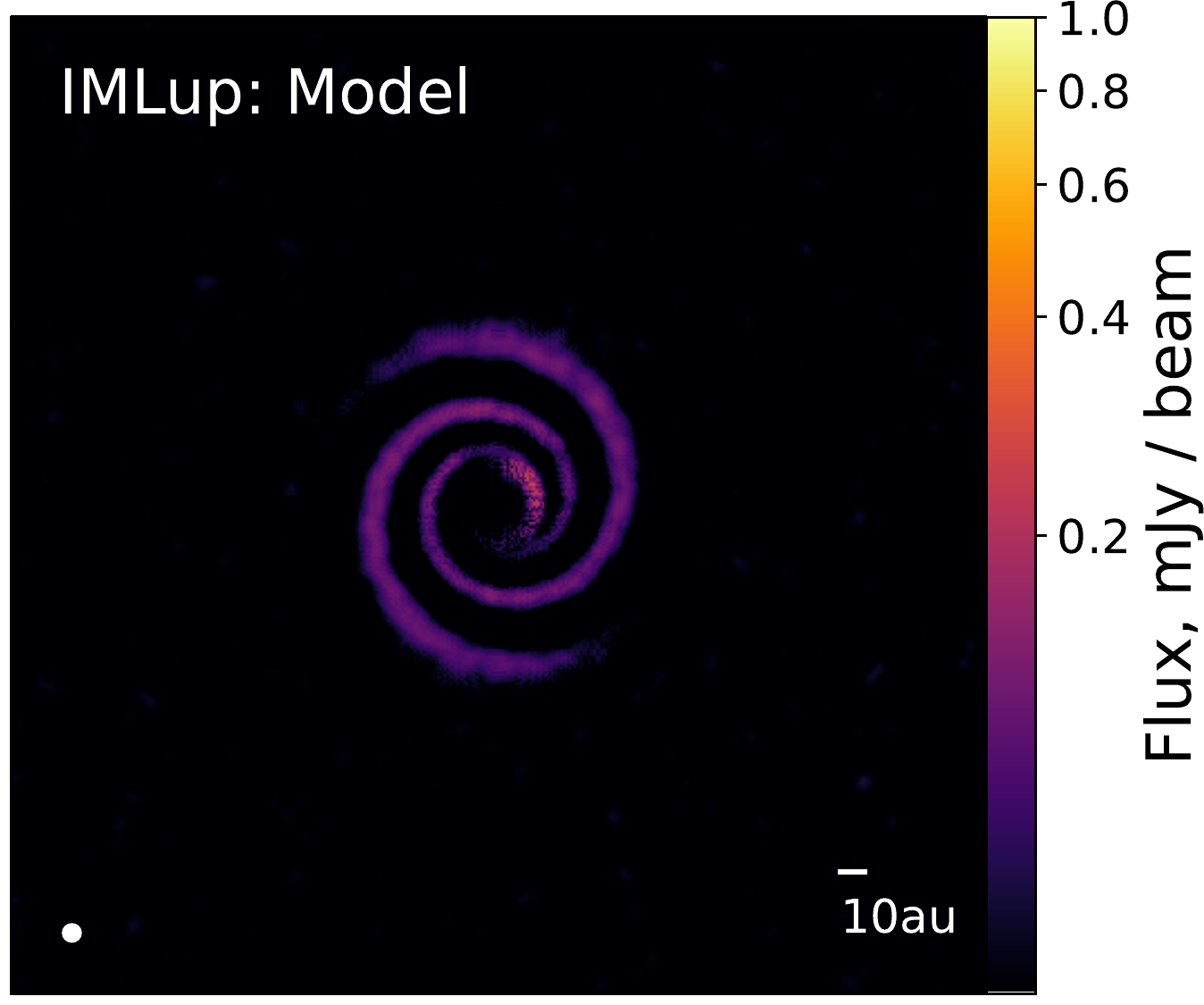}
    \caption{\label{fig:imlup} IM Lup discs images. Top: Deprojected ALMA continuum image (left) and residual profile (right). Bottom: Deprojected disc model continuum image (left) and residual profile (right). Input properties for the disc models and observation parameters are laid out in Tables \ref{tab:dsharp} and \ref{tab:dsharpcasa}.}
\end{figure}

We calculate IM Lup to have a disc mass $M_{\rm disc}=0.098$\,M$_{\odot}$ and $q=0.11$ within $R=110$\,AU, and therefore the lowest disc-to-star mass ratio of the three discs modelled here. The deprojected disc images in Figure \ref{fig:imlup} show tightly wound spiral structure in the continuum and residual images, with geometry and spiral fluxes closely matching those observed in the inner disc of the IM Lup system.

\subsection{Conclusions on DSHARP sample}

We present the results of our semi-analytic analysis of the observed spiral structure in the Elias 27, WaOph 6 and IM Lup systems. We note again that the purpose of this simple functional formalism is not to exactly reproduce, but to approximate, the likely spiral morphologies and fluxes of the 3 systems in question, and to investigate whether systems of their quoted disc and stellar properties should be capable of generating detectable non-axisymmetric substructure when observed with ALMA. We do this by imposing logarithmic spiral structure characteristic of GI, with self-consistently calculated amplitudes and realistic grain distributions. All 3 of the models presented here produce detectable spirals of comparable structure and fluxes to their observed DSHARP counterparts, indicating that GI may be the dominant mechanism responsible for the observed substructure in these discs.

For Elias 27, WaOph 6 and IM Lup we derive disc masses $M_{\rm disc}=0.13$\,M$_{\odot}, 0.16$\,M$_{\odot}$ and $0.098$\,M$_{\odot}$ and disc-to-star mass ratios $q=0.27, 0.24$ and $0.11$ within their respective outer radii. Common assumption is that GI requires $q\gtrsim0.5$, therefore rendering these discs too low mass to generate prominent self-gravitating structure. However it may be possible for discs to display self-gravitating spirals for much lower mass ratios than previously thought, with the critical mass ratio having a strong dependence on the host star mass and disc opacity \citep{veronesietal19,cadmanetal20,haworthetal20}. We therefore should caution against discarding GI as a plausible mechanism based off this simple mass criterion alone.

It is important to note here that whilst we shouldn't be surprised that our models accurately reproduce the spiral form of the systems considered here, as the geometry is imposed in equation \ref{eq:spiraltheta}, we should be more concerned with how accurately our models are able the reproduce the spiral flux amplitudes of the observed systems, as these are determined self-consistently from the disc mass accretion rate and the viscous-$\alpha$. The self-consistently calculated spiral amplitudes in our models all generate comparable fluxes to their counterpart observations, indicating that self-gravity may be a plausible explanation to these 3 systems.

In our model we assume that some grain growth has occurred up to the fragmentation threshold. We note again here that models of grain growth generally suggest that centimeter aggregates form rapidly on timescales $\sim10^5$\,yrs \citep{dullemondetal05,laibetal08}, therefore given the ages of these 3 systems our assumption seems reasonable. If, however, we modelled these systems assuming no grain growth, it is likely that we would not find any signatures of GI. Therefore, if these discs are indeed gravitationally unstable, our models suggest that significant grain growth must have also occurred. Future multi-wavelength observations of these systems, and derivation of the discs' $\beta$-parameter (equation \ref{eq:beta}), will help to establish if this is the case.

An alternative explanation for the observed spiral structure in the DSHARP discs may be the presence of a stellar or planetary-mass companion. Planet-disc gravitational interactions can generate disc perturbations, and massive companions may be capable of triggering two-armed symmetric spiral responses similar to those observed in DSHARP \citep[e.g.][]{dongetal15b,dongetal16,baezhu2018a,baezhu2018b,dsharp4}. However in order to drive the spiral modes observed, for example in the Elias 27 system, would require a wide-orbit companion of potentially tens of Jupiter masses, thus rendering any companion likely detectable at sub-mm/IR wavelengths \citep{meruetal17}. To our knowledge no companion has as yet been detected in any of the 3 discs observed here. More commonly associated features of planet-disc interactions are the presence of annular substructures such as rings and planet-driven gaps. Elias 27, WaOph 6 and IM Lup all display these features, as do a total of 18 discs in the DSHARP sample \citep{dsharp2}. The DSHARP collaboration report no companion detections in any of these 18 discs despite many of the observed features being suggestive of massive companions which ought to be observable at such high angular resolution. It may then be the case that either massive planets are fainter than previously thought \citep{dongetal18}, or that the observed rings are driven by lower mass, fainter planets which remain invisible to the DSHARP survey. If the latter, then these lower mass companions may not be capable of driving the observed spiral structure in Elias 27, WaOph 6 and IM Lup alone, but a combination of both GI and planet-disc interactions may be a plausible scenario \cite[e.g.][]{perezetal16}.

More detailed analysis of these systems, investigating the effect of varying accretion rate, disc irradiation, the dominant spiral mode and grain size distribution will be the subject of future work.

\section{Summary and Conclusion}\label{sec:discussion}

We present our updated self-consistent, semi-analytic model of self-gravitating discs that also includes a prescription for dust trapping. We capitalise on the efficient nature of the model by generating a suite of disc models at little computational expense, and examine the parameter space within which we predict self-gravitating discs will generate spiral structure that can be resolved when imaged with ALMA. Monte-Carlo radiative transfer is employed here to produce synthetic observations of these model discs, allowing us to make realistic predictions about the strength of the perturbations and the grain size distribution required to generate observable spiral structure.

Realistic dust trapping is modelled using a semi-analytic prescription in which particles with $\rm St=1$ may reach grain concentration factor $\eta \approx 6$ at the density peaks of the spiral perturbations, where $\eta$ represents the local dust enhancement relative to the mean dust-to-gas ratio in the disc, assumed to be 0.01 in all the models considered here. We find that particles of millimetre and centimetre sizes concentrate most strongly in spiral arms resulting in significantly enhanced millimetre emission in these regions. When the dust mass budget is dominated by these millimetre and centimetre sized grains we find self-gravitating structure to be observable in much lower mass discs than previously predicted. Through calculation of the grain fragmentation threshold in the discs modelled here we find that grains may only grow to as large as a few centimetres before grain-grain collisions become destructive. Therefore it may be the case that grain size distributions in self-gravitating discs satisfy this dust mass budget criterion.

Our synthetic unsharp masked images of discs in the Taurus star forming region ($d\sim140$\,pc) exhibit distinguishable spiral structure for disc masses as low as $q=0.1$ given sufficient grain growth. These images are generated using realistic ALMA observing setups with reasonable observing times and PWV levels. We do however note that we only consider face-on discs during this evaluation and that inclining and rotating them may well obscure any substructure, likely most adversely in low mass discs with the weakest spirals.

Through multi-wavelength observations and derivation of the $\beta-$parameter we show how it is possible to retrieve information about grain growth and the dust-to-gas ratio distribution from our model discs. Through comparison of our predicted $\beta-$values to those calculated from future multi-wavelength observations of self-gravitating discs, it may be possible to utilise our disc model to examine grain distributions in the observed discs.

Applying our disc model to systems from the DSHARP sample, we find the quoted disc parameters for Elias 27, WaOph 6 and IM Lup suggest that they are all capable of driving observable, self-gravitating spiral structure providing that grains have grown to as large as the fragmentation threshold. We calculate disc-to-star mass ratios $q=0.27, 0.24$ and $0.11$, within their published outer radii, respectively for the 3 systems. A more detailed analysis exploring the potential parameter space of the DSHARP sample will be left to future work.

\section*{Acknowledgements}
We thank the anonymous referee for their insightful  comments which have improved the clarity of this paper. CH is a Winton Fellow and this work has been supported by Winton Philanthropies / The David and Claudia Harding Foundation. This work used the Cirrus UK National Tier-2 HPC service at EPCC (http://www.cirrus.ac.uk) funded by The University of Edinburgh and EPSRC (EP/P020267/1).
%%%%%%%%%%%%%%%%%%%%%%%%%%%%%%%%%%%%%%%%%%%%%%%%%%

%%%%%%%%%%%%%%%%%%%% REFERENCES %%%%%%%%%%%%%%%%%%

\bibliographystyle{mnras}
\bibliography{main} 

%%%%%%%%%%%%%%%%%%%%%%%%%%%%%%%%%%%%%%%%%%%%%%%%%%

\appendix

\section{Gallery of Discs}\label{appendix:gallery}

\begin{figure*}
    \includegraphics[width=.9\linewidth]{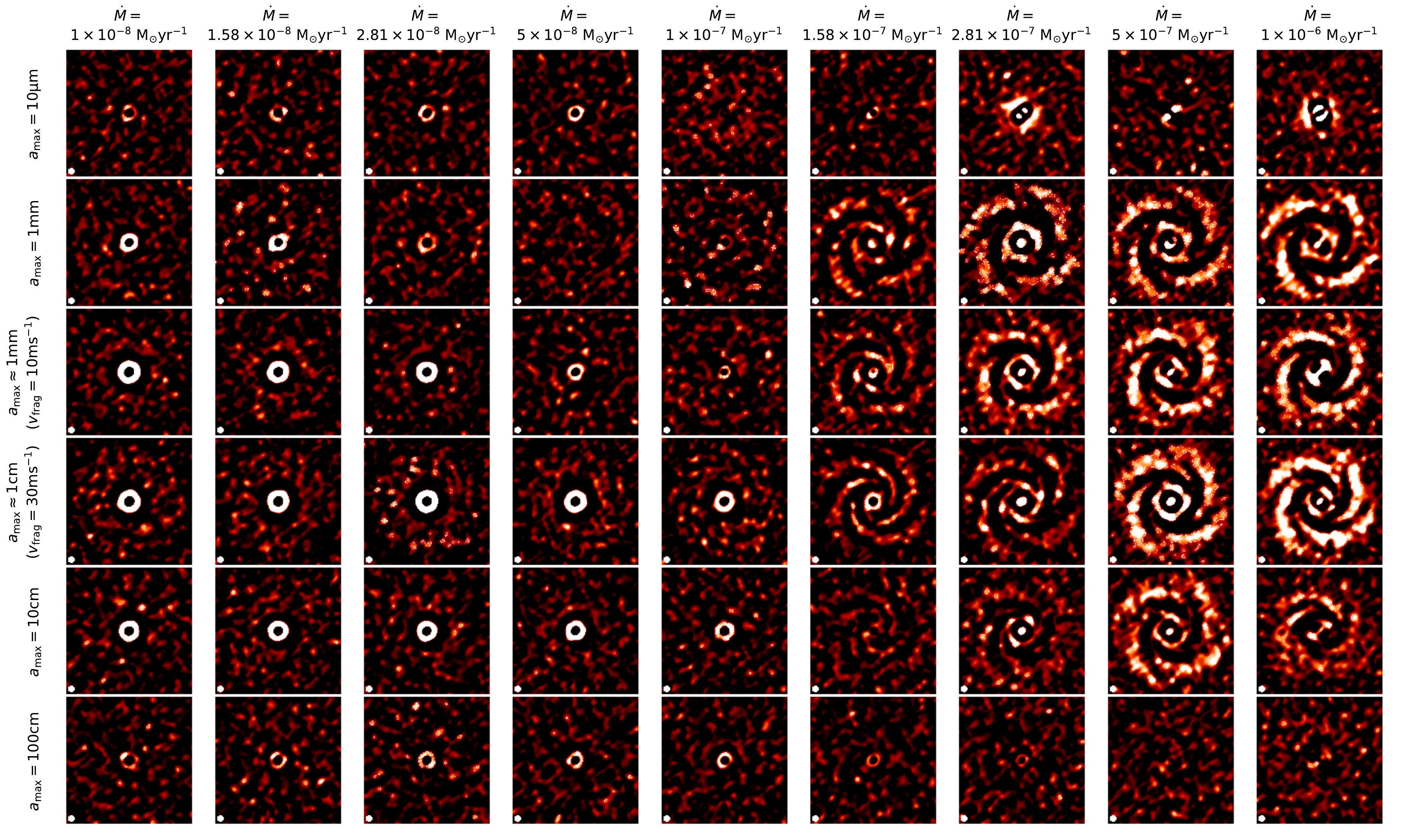}
    \raisebox{0.5\height}{\includegraphics[height=50mm]{cbar_115GHz.png}}
    \caption{\label{fig:115gallery} Gallery of discs observed at $f_{\rm obs}=115$GHz ($\lambda=2.6$\,mm). Disc setups are described in Section \ref{sec:discparams}. \textsc{casa} observing inputs are laid out in Table \ref{tab:casa}.}
\end{figure*}
\begin{figure*}
    \includegraphics[width=.9\linewidth]{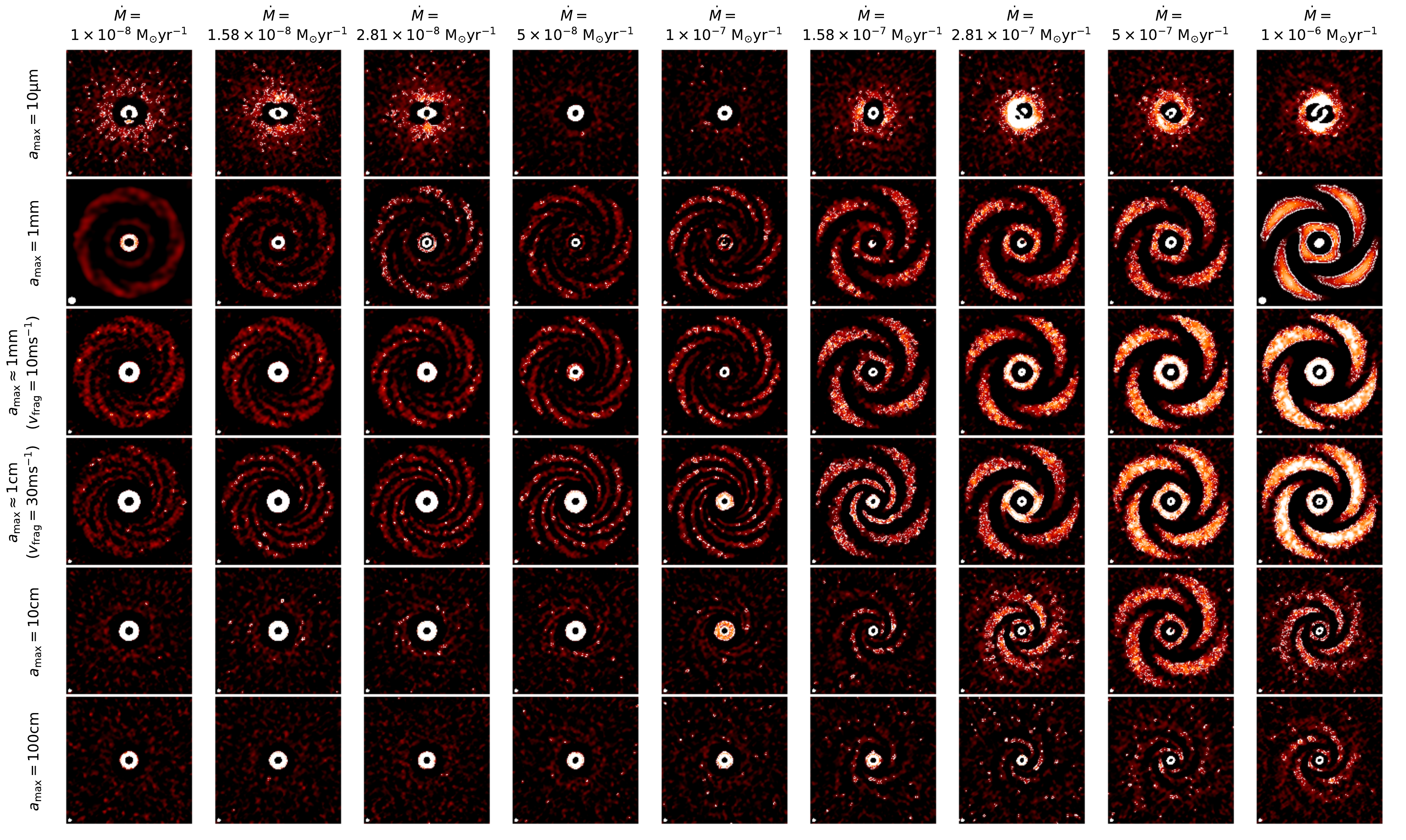}
    \raisebox{0.5\height}{\includegraphics[height=50mm]{cbar_230GHz.png}}
    \caption{\label{fig:230gallery} Gallery of discs observed at $f_{\rm obs}=230$GHz ($\lambda=1.3$\,mm). Disc setups are described in Section \ref{sec:discparams}. \textsc{casa} observing inputs are laid out in Table \ref{tab:casa}.}
\end{figure*}
\begin{figure*}
    \includegraphics[width=.9\linewidth]{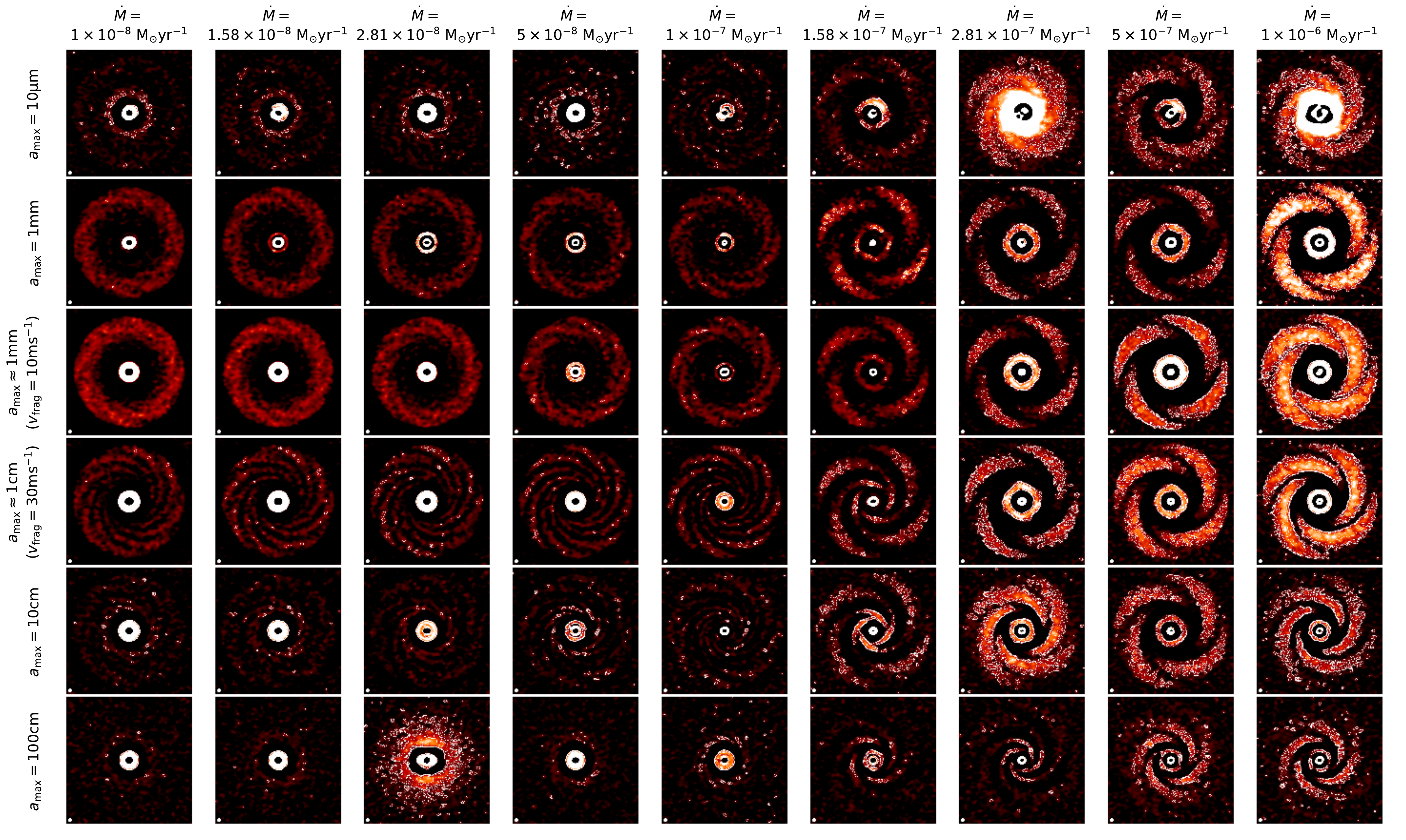}
    \raisebox{0.5\height}{\includegraphics[height=50mm]{cbar_690GHz.png}}
    \caption{\label{fig:690gallery} Gallery of discs observed at $f_{\rm obs}=690$GHz ($\lambda=0.4$\,mm). Disc setups are described in Section \ref{sec:discparams}. \textsc{casa} observing inputs are laid out in Table \ref{tab:casa}.}
\end{figure*}

% Don't change these lines
\bsp	% typesetting comment
\label{lastpage}
\end{document}